\documentclass[aps,prx,reprint,groupedaddress]{revtex4-2}

\usepackage{graphicx}
\usepackage[caption=false]{subfig}
\usepackage{amssymb}
\usepackage{amsmath}
\usepackage{float}
\usepackage{dcolumn}
\usepackage{physics}
\usepackage{bm}
\usepackage{bbold}
\usepackage{tikz}
\graphicspath{ {figures/} }
\usepackage{placeins}
\usepackage{comment}

\newcommand{\be}{\begin{equation}}
\newcommand{\ee}{\end{equation}}
\newcommand{\bea}{\begin{eqnarray}}
\newcommand{\eea}{\end{eqnarray}}
\newcommand{\Eq}[1]{Eq.\,(\ref{#1})}
\newcommand{\Eqs}[2]{Eqs.\,(\ref{#1}) and (\ref{#2})}
\newcommand{\Eqsss}[3]{Eqs.\,(\ref{#1}), (\ref{#2}), and (\ref{#3})}

\newcommand{\Fig}[1]{Fig.\,\ref{#1}}
\newcommand{\Figs}[2]{Figs.\,\ref{#1} and \ref{#2}}

\newcommand{\Sec}[1]{Sec.\,\ref{#1}}
\newcommand{\Onlinecite}[1]{Ref.\,\onlinecite{#1}} 
\newcommand{\App}[1]{Appendix\,\ref{#1}}

\newcommand{\rmand}{\quad{\rm and}\quad}

\begin{document}
	

\title{Controlling dephasing of coupled qubits via shared-bath coherence }
	
\author{L. M. J. Hall}
\email{Luke.Hall415@gmail.com}

\author{L. S. Sirkina}%

\author{A. Morreau}

\author{W. Langbein}

\author{E. A. Muljarov}
\email{egor.muljarov@astro.cf.ac.uk}	
	
\affiliation{%
School of Physics and Astronomy, Cardiff University, Cardiff CF24 3AA, United Kingdom
}
	
\date{\today}
	
\begin{abstract}

The interaction of a quantum system with its environment limits its coherence time. This, in particular, restricts the utility of qubits in quantum information processing applications. In this paper, we show
that the decoherence of a coupled qubit system can be minimized, or even eliminated, by exploiting the quantum coherence of the bath itself. We investigate the dephasing in a system of two spatially separated, electronically decoupled qubits, with direct or mediated coupling, interacting with a shared bath. For illustration, we treat F\"orster or cavity-mediated coupling between semiconductor quantum dots interacting with acoustic phonons. Using the rigorous method of Trotter's decomposition with cumulant expansion, we demonstrate a reduction in the dephasing rates at specific distances between the dots. The control of dephasing with distance is a coherent effect of the shared bath and is absent for independent baths. It can be understood in terms of phonon-assisted transitions between the entangled qubit states of the coupled system.

\end{abstract}

\maketitle

\section{Introduction}
A quantum bit, or qubit, is a two-level quantum-mechanical system. While in many ways, the qubit is analogous to the classical binary bit, quantum computing infrastructure is unique in its reliance on coherent superposition of one or more qubits. Two-qubit logic gates, in particular, are a fundamental building block in any quantum computing architecture~\cite{MajerNat07, ImamogluPRL99}. Such gates require a controlled long-range interaction between isolated qubits, which can be mediated by their strong coupling to a photonic cavity~\cite{DelbecqNatCom13}. The lifetime of this interaction, known as the coherence time, dictates the complexity of calculations that can be achieved, and the accuracy of the calculated results. Inevitably, the coupling of the qubits to their environment, often treated as a thermal bath, limits coherence times and hence restricts the practical application of multi-qubit logic gates~\cite{Palma96,SchlosshauerPR19}.

Historically, the dominating source of decoherence in a multi-qubit system was the cavity itself, and consequently the cavity was regarded as the bath. Therefore, previous works have focused on exploiting bath coherent properties to reduce dephasing, such as decoherence-free subspaces of subradiant quantum superpositions~\cite{ZanardiPRL97,LidarPRL98}. Specifically, the introduction of a second qubit coupled to the same cavity gives rise to a subradiant superposition state that is decoupled from the lossy cavity. Although the quality factor of optical cavities dramatically increased over the past decade, coherence times remain limited due to other facets of the environment, the details of which are specific to the physical implementation of the qubit system.

While there are many possible physical implementations of a qubit, we will focus here on semiconductor QDs, often referred to as ``artificial atoms''. They are a promising qubit candidate, since quantum interference of single photons emitted by spatially separated GaAs QDs has been experimentally demonstrated, highlighting the underlying coherence required between these photon sources to achieve such interference~\cite{FlaggPRL10,PatelNatPhot10,ZhaiNatNano22}. Coupling these QDs to optical cavities further enhances this effect~\cite{GieszPRB15}, and maintaining this coherence is of great importance for applications in quantum computing. However, in semiconductor QDs, acoustic phonons present the major intrinsic source of decoherence. Even at low temperatures, acoustic phonons induce a rapid non-Markovian decay of the QD coherence~\cite{BorriPRL01,KrummheuerPRB02}, also known as a phonon broad band (BB) in the QD spectrum, followed by a nearly Markovian long-time decay of the zero-phonon line (ZPL) due to real or virtual phonon-assisted transitions to other QD levels~\cite{MuljarovPRL04,MuljarovPRL05}.

The QD interaction with a phonon bath fundamentally differs from the bilinear QD-cavity coupling, so that the aforementioned idea of decoherence-free subspaces of qubit states is not directly applicable here. Nevertheless, progress has been made to reduce the effect of QD decoherence in qubit control. In particular, using a controlled off-resonant optical pulse with the laser pulse frequency tuned to the BB allows one to prepare almost pure qubit states by using phonon assisted transitions~\cite{GlasslPRL13,QuilterPRL15}. Notably, this only applies in the low temperature regime, where phonon absorption can be reasonably neglected.  The idea has been generalized to a phonon-assisted two-photon excitation scheme to create indistinguishable entangled photon pairs from remote QDs~\cite{ReindlNL17}. The Purcell effect helps to reduce the phonon-induced decoherence by a resonant weak coupling of a QD exciton to a cavity mode that results in reduction of the relative weight of the BB and enhancement of the ZPL emission~\cite{GrangePRL17}. Moreover, in the QD-cavity strong coupling regime, the BB is almost entirely eliminated in the cavity excitation scheme~\cite{MorreauArxiv,SirkinaPRB23}. However, the ZPL gains an additional dephasing~\cite{Wilson-RaePRB02} which can be understood and quantified in terms of phonon-assisted transitions between the polariton states of the system~\cite{MorreauPRB19}. Such a ZPL dephasing can be enhanced in coupled qubits where all parts of the system are interacting with the bath.

In this paper, we demonstrate a reduction, or even a complete elimination, of the ZPL dephasing in a system of two QD qubits coupled to each other directly or via an optical cavity and interacting with a bath of acoustic phonons. We show that, while the interaction of entangled qubits with a shared environment usually causes dephasing of qubit states, the coherent properties of the bath can help to reduce this decoherence, which also improves the gate fidelity.

\section{System Hamiltonian}

As a practical example, we consider the decoherence of electronically decoupled qubits separated by a distance $d$ and interacting with a shared bath. The coupling of the qubits is taken as either direct through dipolar F\"orster-type coupling~\cite{GovorovPRB05,RozbickiPRL08,ThilagamJPCM08,GribbenPRR20}, or indirect by cavity-mediation~\cite{ReitzensteinOL06,AlbertNatCom13,WoerkomPRX18}, or both. As qubit and bath realisation we use  semiconductor QDs interacting with a bath of three-dimensional (3D) acoustic phonons, widely studied in the literature~\cite{BorriPRL01,KrummheuerPRB02,MuljarovPRL04,MuljarovPRL05,Grosse08}.

\begin{figure}[t]
\centering
\includegraphics[width=0.3\textwidth]{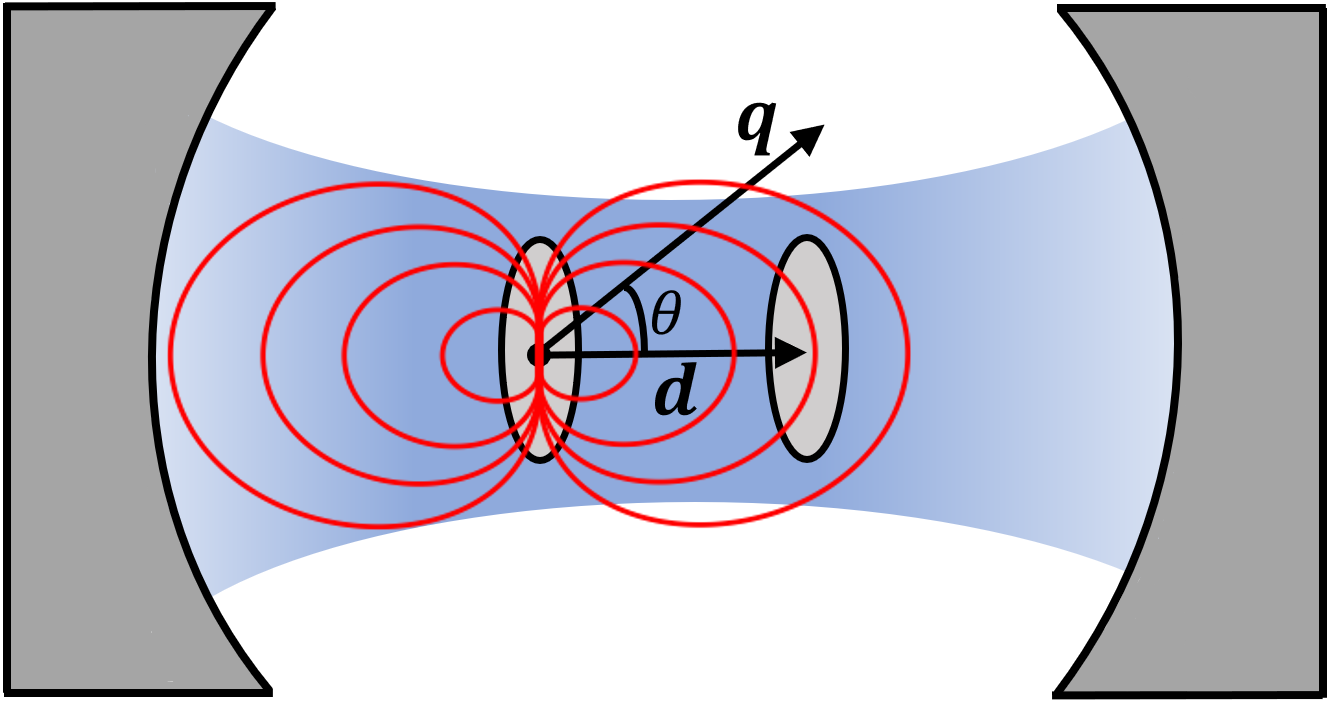}
\caption{Schematic of the system with a pair of dipole-coupled qubits separated by a distance vector $\textbf{d}$, coupled to an optical cavity, and interacting with three-dimensional acoustic phonons described by a wave vector $\textbf{q}$ and angle $\theta$.}
\label{schematic}
\end{figure}

The system Hamiltonian can be written as a sum of two exactly solvable parts,
\begin{equation}
H = H_{0} + H_\text{\rm IB}\,,
\label{Ham}
\end{equation}
where $H_0$ describes the coupling between the qubits and the cavity, and $H_\text{\rm IB}$ is a generalized independent boson (IB) model Hamiltonian describing the coupling of the qubits to the shared environment. For the system of two remote QDs coupled to an optical cavity, illustrated in \Fig{schematic}, $H_0$ takes the form (with $\hbar=1$)
\begin{equation}\label{H0}
\begin{split}
H_{0} =& \; \Omega_1 d_1^\dagger d_1 + \Omega_2 d_2^\dagger d_2 + \Omega_C a^\dagger a + g (d_1^\dagger d_2 + d_2^\dagger d_1) \\ &+ g_1 (d_1^\dagger a + a^\dagger d_1) + g_2 (d_2^\dagger a + a^\dagger d_2)\,,
\end{split}
\end{equation}
where  $d_{j}^\dagger$ is the fermionic exciton creation operator in QD $j$ ($j=1,2$), $a^\dagger$ is the cavity photon creation operator, $\Omega_{j}$ ($\Omega_{C}$) is the exciton (cavity photon) frequency, and $g$ and $g_{j}$ are the coupling strengths between the QD excitons, and the exciton in QD $j$ and the cavity photon, respectively. The IB model Hamiltonian describes the interaction of the QD excitons with a shared acoustic-phonon bath,
\begin{equation}\label{HIB}
H_\text{\rm IB} = H_\text{\rm ph} + d_1^\dagger d_1 V_1 + d_2^\dagger d_2 V_2\,,
\end{equation}
where
\begin{equation}\label{Hph_V}
H_\text{\rm ph}=\sum_{\textbf{q}} \omega_q b_{\textbf{q}}^\dagger b_{\textbf{q}}\ \ \ {\rm and} \ \ \
V_{j} = \sum_{\textbf{q}} \lambda_{\textbf{q},j} (b_{\textbf{q}} + b_{-\textbf{q}}^\dagger)
\end{equation}
are, respectively, the free phonon bath Hamiltonian and the QD coupling to the bath, where $b_{\textbf{q}}^\dagger$ is the bosonic creation operator of a bulk phonon mode with the momentum $\textbf{q}$ and frequency $\omega_q$ (denoting $q = \lvert \textbf{q} \rvert$). The coupling of the exciton in QD $j$ to the phonon mode $\textbf{q}$ is given by the matrix element $\lambda_{\textbf{q},j}$,
which depends on the material parameters, exciton wave function, and position of the QD. Their explicit form for isotropic and anisotropic QDs is provided in \App{App:Coupling}. For identical QD qubits separated by a distance vector $\textbf{d}$, the matrix elements satisfy the relation
\begin{equation}\label{lambda}
\lambda_{\textbf{q},2} = e^{i\textbf{q}\cdot \textbf{d}}\lambda_{\textbf{q},1}\,.
\end{equation}

\section{Asymptotically exact solution}

In the following, we focus on the linear optical polarization, which allows us to study the coherence of the system as a function of the distance between the qubits. The linear polarization of qubit $j$ is defined as
$P_{jk}(t)={\rm Tr}\{\rho(t) d_j\}$, where $\rho(t)$ is the full density matrix. We assume that starting from the system ground state the qubit with index $k$ is instantaneously excited at time $t=0$. As has been derived in Ref.~\cite{MorreauPRB19}, the linear polarization can written as
\begin{equation}\label{LinPol}
P_{jk}(t) = \langle \bra{j} \hat{U}(t) \ket{k} \rangle_\text{\rm ph}\,,
\end{equation}
where $\hat{U}(t) = e^{iH_\text{\rm ph}t} e^{-iHt}$ is the evolution operator and $\langle...\rangle_\text{\rm ph}$ denotes the expectation value over all phonon degrees of freedom in thermal equilibrium. Here and below we use the basis states
\begin{equation}
\ket{j}=d_j^\dagger \ket{0} \rmand \ket{C}= a^\dagger \ket{0}\,,
\end{equation}
where $\ket{0} $ represents the vacuum state of the QD-cavity subsystem, and $j=1,2$.

Taking advantage of the two exactly solvable parts of the Hamiltonian \Eq{Ham}, we apply the method of Trotter's decomposition with cumulant expansion~\cite{MorreauPRB19} summarized in the following section, allowing to take into account the effect of the phonon environment {\it exactly}.

\subsection{Trotter's decomposition}

The method of Trotter's decomposition with linked cluster expansion developed in \cite{MorreauPRB19} for exact calculation of the linear polarization of a single QD simultaneously coupled to a cavity and a phonon bath is applied here to the more general case of cavity-mediated coupling between the QDs (with the coupling constants $g_1$ and $g_2$) and their direct dipolar coupling (with the coupling constant $g$), as described by \Eq{H0}. We commence by splitting the time interval $[0,t]$, where $t$ is the observation time, into $N$ equal steps of duration $\Delta t = t/N = t_n - t_{n-1}$, where the time $t_n = n\Delta t$ represents the time after the $n$-th step. Trotter's theorem is then used to separate the time evolution of the two non-commuting operators, ${H}_0$ and $H_\text{\rm IB}$. For sufficiently small $\Delta t$, we can assume independent evolution of the two exactly solvable components within each time step. In fact, applying Trotter's decomposition theorem, the time evolution operator $\hat{U}(t)$ can be written as
\begin{equation}\label{TrottersU}
\hat{U}(t) = \lim_{N \to \infty} e^{iH_\text{\rm ph}t}(e^{-iH_\text{\rm IB} t/N} e^{-iH_0 t/N})^N\,.
\end{equation}
We now introduce two operators $\hat{M}$ and $\hat{W}$, which describe the dynamics due to $H_0$ and $H_\text{\rm IB}$, respectively, each being analytically solvable. Using these operators, the QD-cavity dynamics over a single time step is described by
\begin{equation}\label{MoperatorForster}
\hat{M}(t_n-t_{n-1}) = \hat{M}(\Delta t) = e^{-iH_0\Delta t}
\end{equation}
and the exciton-phonon dynamics is given by
\begin{equation}\label{WoperatorForster}
\hat{W}(t_n,t_{n-1}) = e^{iH_\text{\rm ph}t_n}e^{-iH_\text{\rm IB}\Delta t} e^{-iH_\text{\rm ph}t_{n-1}}\,.
\end{equation}
Exploiting the commutativity of $H_0$ and $H_\text{\rm ph}$, one can write the time evolution operator \Eq{TrottersU} as
\begin{equation}\label{UasproductWM}
\hat{U}(t) = \mathcal{T} \prod_{n=1}^N \hat{W}(t_n,t_{n-1})\hat{M}(t_n -t_{n-1})\,,
\end{equation}
where $\mathcal{T}$ is the time-ordering operator.
$\hat{W}$ and $\hat{M}$ are both $3 \times 3$ matrices in the $\ket{1}$, $\ket{2}$, $\ket{C}$ basis, and due to the diagonal form of the exciton-phonon interaction, $\hat{W}$ is diagonal. Its diagonal matrix elements can be written as
\begin{equation}
W_{i_n} (t_n,t_{n-1}) =  \mathcal{T} \exp{ -i \int_{t_{n-1}}^{t_n} \tilde{V}_{i_n}(\tau) d\tau }\,,
	\end{equation}
where
\begin{equation}\label{Vbetanu}
\tilde{V}_{i_n}(\tau) = \xi_{i_n} {V}_1(\tau) + \eta_{i_n} {V}_2(\tau)
\end{equation}
for $\tau$ within the time interval $t_{n-1} \leqslant \tau \leqslant t_{n}$\,, with  $\xi_{i}$ and $\eta_{i}$ being the components of the vectors $ \vec{\xi} = (1,0,0)$ and $\vec{\eta} = (0,1,0)$,
respectively, and $V_j(\tau) = e^{iH_\text{\rm ph}\tau}V_je^{-iH_\text{\rm ph}\tau}$ is the exciton-phonon coupling in the interaction representation, with $V_j$ defined in \Eq{Hph_V}. We use the indices $i_n$ to indicate which state the system takes at a given time step $n$, being either $\ket{1}$, $\ket{2}$, or $\ket{C}$, with $i_n$ taking the values 1, 2, or $C$, respectively. The elements of $\vec{\xi}$ and $\vec{\eta}$ selected by $i_n$ determine the exciton-phonon coupling used during the $n$-th time step. For example, if the system is in the first QD exciton state during the $n$-th time step, then $i_n =1$, and the exciton-phonon interaction $V_1$ occurs.

To find the polarization, we use \Eq{UasproductWM} to substitute $\hat{U}(t)$ in \Eq{LinPol}, and write the matrix products explicitly, yielding
\bea
P_{jk}(t) &=& \sum_{i_{N-1} = 1,2,C} \dots \sum_{i_{1} = 1,2,C} M_{i_N i_{N-1}}
\dots M_{i_1 i_{0}}
\nonumber\\
&&\times\langle W_{i_N}(t,t_{N-1}) \dots W_{i_1}(t_1,0)\rangle_\text{\rm ph}\,,
\label{PolarizationMthenWForster}
\eea
where $i_0 = k$ and $i_N = j$ denote, respectively, the excitation channel $k$ at $t=0$ and measurement channel $j$ at the final time step $t_N=t$, and $M_{i_n i_m}=[\hat{M}(\Delta t)]_{i_n i_m}$. The  $W_{i_n}$ operators include the phonon contributions, therefore we separate this product from the rest of the expression in order to take the expectation value and apply the linked cluster theorem \cite{Mahan00,MuljarovPRL04,MuljarovPRL05}.

\subsection{ Linked cluster expansion}

To calculate the expectation value of the products of the exciton-phonon interaction operators in \Eq{PolarizationMthenWForster}, we apply the linked cluster theorem. It allows us to write this expectation value as an exponential with a double sum over all possible second-order cumulants in the exponent \cite{Mahan00,MorreauPRB19},
\bea\label{tracecumulantForster}
\langle W_{i_N}(t,t_{N-1})\dots  W_{i_1}(t_1,0)\rangle_\text{\rm ph} \nonumber\\
\quad =\exp(\sum_{n=1}^N \sum_{m=1}^N  \mathcal{K}_{i_n i_m}(|n-m|))
\eea
with the cumulants in \Eq{tracecumulantForster} given by
\begin{equation}\label{Knm}
\mathcal{K}_{i_n i_m}(s)  =   -\frac{1}{2} \int_{t_{n-1}}^{t_n}\hspace*{-4mm}d\tau_1 \int_{t_{m-1}}^{t_m} \hspace*{-5mm}d\tau_2
\langle \mathcal{T} \tilde{V}_{i_n}(\tau_1) \tilde{V}_{i_m}(\tau_2)\rangle_\text{\rm ph},
\end{equation}
where $s=|n-m|$. Using \Eq{Vbetanu}, this cumulant can be expressed as
\bea
\mathcal{K}_{i_n i_m}(s) &=&  \xi_{i_n}\xi_{i_m} K_{11}(s) + \eta_{i_n} \eta_{i_m} K_{22}(s)
\nonumber\\
&&+ (\xi_{i_n}\eta_{i_m}+\eta_{i_n}\xi_{i_m})K_{12}(s)\,,
\label{Kinim}
\eea
where we have introduced the cumulant elements
\begin{equation}
\label{Kjj'}
K_{jj'}(s) =  -\frac{1}{2}  \int_{t_{n-1}}^{t_n} d\tau_1 \int_{t_{m-1}}^{t_m} d\tau_2 D_{jj'} (\tau_1 - \tau_2)\,,
\end{equation}
having the symmetry $K_{jj'}(s)=K_{j'j}(s)$. Here we use the phonon Green's function
\begin{equation}
D_{jj'}(t) = \int_0^\infty d\omega \; J_{jj'}(\omega) D(\omega,t)\,,
\end{equation}
in which
\begin{equation}\label{spectraldens}
J_{jj'}(\omega) = \sum_\textbf{q} \lambda_{\textbf{q},j}\lambda_{\textbf{q},j'}^\ast \delta (\omega-\omega_q)
\end{equation}
is the phonon spectral density
and
\begin{equation}
D(\omega,t) = [{\cal N}_\omega +1] e^{-i\omega |t| }  + {\cal N}_\omega  e^{i\omega |t|}
\end{equation}
is the phonon propagator, where  ${\cal N}_\omega=[e^{\omega/T}-1]^{-1}$ is the Bose distribution function and $T$ is the temperature (using units with the Boltzmann constant $k_B=1$). The explicit analytical forms of the spectral density $J_{jj'}(\omega)$ for isotropic and anisotropic QDs are derived in \App{App:Coupling}. The cumulant elements \Eq{Kjj'} can be conveniently expressed as linear combinations of the values on the time grid of the cumulant function
\bea
\label{Cjj}
C_{jj'}(t)&=& -\frac{1}{2}  \int_{0}^{t} d\tau_1 \int_{0}^{t} d\tau_2 \; D_{jj'}(\tau_1 - \tau_2)\, ,
\eea
as detailed in \App{App:Cumulants}.

The linear polarization then takes the form
given by \Eqsss{PolarizationMthenWForster}{tracecumulantForster}{Kinim}.
Note that a particular realization or a path of the system evolution is indicated by the indices $i_1,i_2,\dots i_{N-1}$ in \Eq{PolarizationMthenWForster}. However, to obtain the full quantum dynamics of the system, all possible realizations are to be summed over, in line with the path integral formalism. Technically this means a summation over all of the indices, which is done in \Eq{PolarizationMthenWForster}.

	
\subsection{The $L$-neighbor ($L$N) approach}
\label{Sec-LN}

For a finite bath memory time, it is sufficient to consider only a portion of the grid at least up to the memory time, which is referred to as the number of neighbors $L$, defined as the maximum value of $|n-m|$ taken into account in the calculation. The $L$N approach is used to describe the temporal correlations between all considered steps within the memory kernel.
\begin{figure}[t]
\centering
\includegraphics[scale=0.6]{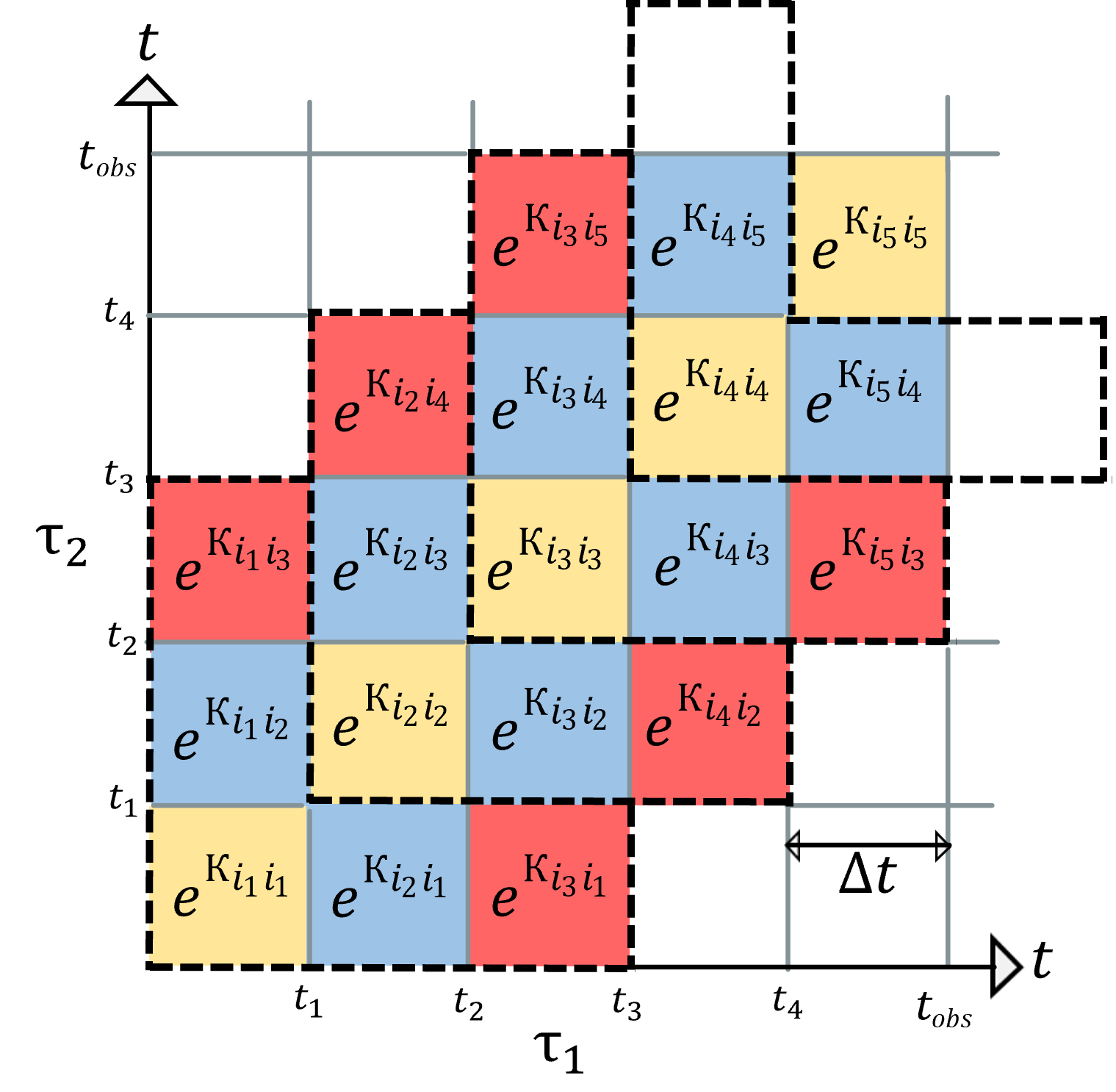}
\caption{\label{timegrid} A portion of the time grid used in the $L$N approach with $L=2$, showing the self interaction (yellow squares), and the nearest (blue squares) and next-nearest neighbor interactions (red squares). }
\end{figure}
We first define a quantity $F_{i_L\ldots i_1}^{(n)}$ which is generated via the recursive relation
\begin{equation}
F_{i_L\ldots i_1}^{(n+1)} = \sum_{l=1,2,C} \mathcal{G}_{i_L\ldots i_1 l} F_{i_{L-1}\ldots i_1 l}^{(n)}\,,
\label{Fn}
\end{equation}
using $F_{i_L\ldots i_1}^{(1)}= M_{i_1 k}$ as the initial value, where $k$ is the excitation channel and  $\hat{M}$ is given by \Eq{MoperatorForster}. $\mathcal{G}$ is known as the influence tensor and is given by
\begin{equation}
\label{GtensorLN}
\mathcal{G}_{i_L \dots i_1 l} = M_{i_1 l}  e^{\mathcal{K}_{ll}(0) + 2\mathcal{K}_{i_1 l}(1) + 2\mathcal{K}_{i_2 l} (2)+ \dots + 2\mathcal{K}_{i_L l}(L)  }
\end{equation}
with a more explicit form provided in \App{App:Cumulants}. The influence tensor $\mathcal{G}$ is a memory kernel containing the information required to propagate the system over a single time step. It includes the path segments connecting the current time interval with the $L$ nearest intervals and to itself which are shown by the $L$ shaped black outlines in \Fig{timegrid}. Each element of the tensor corresponds to a particular path of the system evolution within its memory.
The linear polarization is then given by
\begin{equation}\label{PjkLN}
P_{jk}(t) = e^{\mathcal{K}_{jj}(0)} F_{C\dots C j}^{(N)}\,,
\end{equation}
where $j$ is the measurement state. The indices being placed in the cavity ($C$) state have the result of removing the excess contributions from the $\mathcal{G}$ tensor after the observation time $t$ (see \Fig{timegrid}), as being in the cavity state reduces the cumulant at the corresponding times steps to zero. Equation \eqref{PjkLN} provides an asymptotically ($L\rightarrow \infty$) exact solution for the linear polarization.
In practice, we calculate the linear polarization for a set of finite but sufficiently large $L$, up to $L=30$ in this work and extrapolate the result to $L\to\infty$ (see \App{App:Extrapolation} for details on the extrapolation), in this way approaching the exact solution.
This method can be generalized to other elements of the density matrix, such as the four-wave mixing polarization~\cite{SirkinaPRB23} and the population~\cite{sirkinaInPreparation}.

\subsection{Independent phonon baths}
The case of independent phonon baths can be considered as a simplification to the system, described by Eqs.\,(\ref{Ham})--(\ref{Hph_V}) where the relevant modifications to the system Hamiltonian are applied to the $H_\text{\rm IB}$ term,
\begin{equation}
	H_\text{\rm IB} = H_{\text{\rm ph},1} + H_{\text{\rm ph},2} + d_1^\dagger d_1 V_1 + d_2^\dagger d_2 V_2\,,
\end{equation}
which now describes the interaction of each exciton with its own independent phonon bath, given in a similar way to \Eq{Hph_V} by
\begin{equation}
	H_{\text{\rm ph},j}=\sum_{\textbf{q}} \omega_{{q},j} b_{\textbf{q},j}^\dagger b_{\textbf{q},j}\,,\,
	V_j = \sum_{\textbf{q}} \lambda_{\textbf{q},j} (b_{\textbf{q},j} + b_{-\textbf{q},j}^\dagger)\,.
\end{equation}
Initially this may seem like a complication due to the extra terms. However, the resulting cumulant $\mathcal{K}_{i_n i_m}$  in \Eq{tracecumulantForster} is non-vanishing only when $i_n = i_m$, i.e. the phonon Green's functions $D_{12}$ and $D_{21}$ corresponding to the cross terms vanish,
\begin{equation}
	\langle \mathcal{T} \tilde{V}_1(\tau_1)\tilde{V}_{2}(\tau_2)\rangle  = \langle \mathcal{T} \tilde{V}_2(\tau_1)\tilde{V}_{1}(\tau_2)\rangle = 0\,.
\end{equation}
This is because the phonon operators in $V_1$ commute with those in $V_2$. The result is that the cumulant contains only the diagonal elements $K_{jj}(s)$, giving
\begin{equation}\label{KinimIndependent}
	\mathcal{K}_{i_n i_m} (s) =
	\begin{cases}
		\xi_{i_n}^2 K_{11}(s) + \eta_{i_n}^2 K_{22}(s) & i_n = i_m\,\\
		0 & i_n \neq i_m\,,\\
	\end{cases}
\end{equation}
where $s=|n-m|$.
The linear polarization in this case is calculated using \Eqsss{Fn}{GtensorLN}{PjkLN} with the modified cumulant elements \Eq{KinimIndependent}.

\section{Control of decoherence}

For illustration we consider two cases: {\em Case A} of qubits with direct coupling strength $g$ but without coupling to the cavity ($g_1=g_2=0$) and {\em Case B} of qubits without direct coupling ($g=0$) but with indirect coupling mediated by the cavity via $g_1$ and $g_2$. To elucidate the effect of the shared environment on the system coherence and its dependence on the distance $d=|{\bf d}|$ between the qubits, we assume here that the coupling constants $g_1$, $g_2$, and $g$ are distance independent. We also choose without loss of generality that the first QD is instantaneously excited (e.g. by an ultrashort optical pulse), creating an excitonic polarization with $P_{jk}(0)=\delta_{jk}\delta_{k1}$, where $\delta_{jk}$ is the Kronecker delta.

\subsection{Directly coupled QD qubits}

\begin{figure}[t]
	\centering
	\begin{tikzpicture}
		\node[anchor=south west, inner sep=0] (image) at (0,0) {\includegraphics[width=0.45\textwidth]{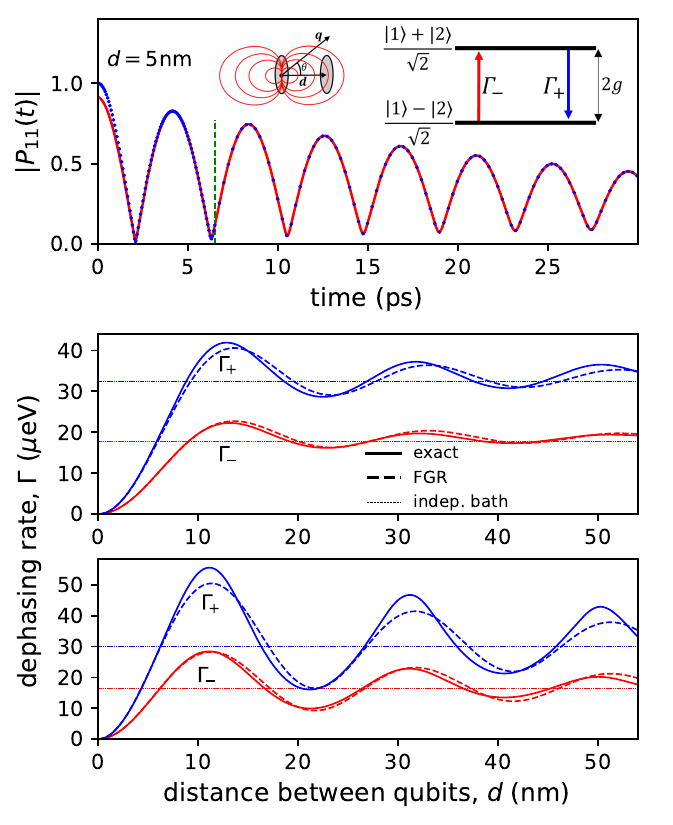}};
		\begin{scope}[x={(image.south east)},y={(image.north west)}]
			\node[anchor=center,font=\sffamily\bfseries] at (0.92,0.955) {(a)};
			\node[anchor=center,font=\sffamily\bfseries] at (0.92,0.576) {(b)};
			\node[anchor=center,font=\sffamily\bfseries] at (0.92,0.298) {(c)};
		\end{scope}
	\end{tikzpicture}
	\caption{(a) Linear optical polarization $P_{11}(t)$ (blue dots) and its complex bi-exponential fit (red lines) for dipolar coupled ($g=0.5$\,meV) isotropic QD qubits (left inset) at zero detuning, separated by the distance $d=5\,$nm, with excitation and measurement in QD $1$. Right inset: energy level diagram for the mixed qubit states, with real phonon-assisted transitions (red and blue arrows). (b,c) Dephasing rates $\Gamma_\pm$ of the mixed states $|\pm\rangle$ as a function of $d$, calculated exactly (solid lines) and via FGR (dashed lines) for (b) isotropic QDs with a confinement length of $l=5.6\,$nm and (c) anisotropic QDs with $l=7.5\,$nm across and $l_\perp=2.5\,$nm along the separation (see \App{iso_QDs} and \App{Aniso_QDs} for details of isotropic and anisotropic QD models, respectively). The rates for independent phonon baths are shown by thin dashed lines. The phonon bath parameters are given in the main text.}
	
	\label{QDQD_Linebroadening}		
\end{figure}

\subsubsection{Linear polarization and dephasing rates}

In {\em Case A}, the time evolution of $P_{11}(t)$ for a system of two dipolar-coupled ($g=0.5$\,meV) identical isotropic QDs of confinement length $l=5.6\,$nm separated by the center-to-center distance $d=5\,$nm, is shown in \Fig{QDQD_Linebroadening}(a) by a blue dotted line, exhibiting decay and oscillations. The phonon bath parameters are taken as $D_c - D_v =-6.5\,$eV, where $D_c$ ($D_v$) is the conduction (valence) band deformation potential, $v_s=4.6\times 10^3\,$m/s is sound velocity, $\rho_m = 5.65\,\text{g}/\text{cm}^3$ is the mass density~\cite{MuljarovPRL04,MuljarovPRL05}, and the temperature is $T=20\,$K.

The behavior in \Fig{QDQD_Linebroadening}(a) is qualitatively explained by the energy level diagram in the right inset, showing hybridized states $|\pm\rangle=(|1\rangle\pm|2\rangle)/\sqrt{2}$ of the two-qubit coupled system at zero detuning ($\Omega_1 = \Omega_2$), where $|1\rangle$ and $|2\rangle$ are the individual QD excited states.  The energy levels are separated by the Rabi splitting $2g$ determining the beat frequency in $|P_{11}(t)|$ which physically expresses the quantum information exchange between the qubits.
The temporal decay of the linear polarization expresses the decoherence in this two-qubit system as a consequence of the interaction of the qubits with the bath. For these QD qubits, the decoherence is due to phonon-assisted transitions between the hybridized states.

With this picture in mind, we have applied to the long-time dynamics of ${P}_{11}(t)$ a biexponential fit of the form
\be
{P}_{11}^{\rm fit}(t)=\sum_j {C}_j e^{ -i\omega_j t }\,,
\ee
extracting the complex amplitudes ${C}_j$, energies $\Re\,\omega_j$, and dephasing rates $\Gamma_j=-\Im\, \omega_j$ of the phonon-dressed mixed states. The fit, applied after the phonon-memory cut-off [introduced in \App{App:Cutoff} by analyzing the cumulant functions and shown in \Fig{QDQD_Linebroadening}(a) by the vertical dashed green line], demonstrates a remarkable agreement with the full calculation with a relative error below $10^{-10}$. At earlier times the deviation is due to the formation of a polaron cloud around the optically excited QD, which is responsible for non-Markovian dephasing and the BB~\cite{KrummheuerPRB02,MorreauPRB19,SirkinaPRB23}. The dephasing rates $\Gamma_j$ extracted from the fit as functions of the QD separation $d$ are shown by solid lines in \Fig{QDQD_Linebroadening}(b) for isotropic and in \Fig{QDQD_Linebroadening}(c) for anisotropic QDs. They are the dephasing rates of the states $|\pm\rangle$, denoted by $\Gamma_{\pm}$, and can be understood as being due to phonon-assisted transitions between the states. At short distances we observe a dramatic increase of the dephasing rates from zero at zero distance (which however cannot be practically realised due to the finite extension of the QDs), followed by an oscillatory behaviour at larger distances. Importantly, the minima of these dephasing rates are lower than the independent baths rates (thin horizontal lines), and are thus a manifestation of shared bath coherence.

\subsubsection{Phonon-assisted transitions between entangled qubit states}
\label{Sec:FGR}
To understand the dependence on the distance between the qubits,
we introduce the fermionic operators
\be
p_\pm^\dagger = D_\mp d^\dagger_1 \pm  D_\pm d^\dagger_2\,,
\label{canon}
\ee
creating excitations of the mixed (i.e. entangled) QD qubit states
\be
\label{mixed}
|\pm\rangle= D_\mp |1\rangle \pm D_\pm |2\rangle\,,
\ee
where
\be
D_\pm = \sqrt{(1 \pm \Delta/R)/2}\,,
\label{Dpm}
\ee
with
\be
\Delta=\Omega_2-\Omega_1 \rmand  R= \sqrt{\Delta^2 + 4g^2}
\ee
being, respectively, the detuning and the Rabi splitting. In the absence of the bath, these operators diagonalize the system Hamiltonian \Eq{H0} exactly:
\be
H_0=\Omega_+ p_+^\dagger p_+ + \Omega_- p_-^\dagger p_-\,,
\ee
where
\be
\Omega_{\pm}= \frac{\Omega_1 + \Omega_2 \pm R} {2}
\label{ompm}
\ee
are the energies of the hybrid states $|\pm\rangle$.

Now applying this canonical transformation to the total Hamiltonian \Eq{Ham} we obtain
\bea
H&=&(\Omega_++ V_+) p_+^\dagger p_+ + (\Omega_- + V_-) p_-^\dagger p_-
\nonumber\\
&&+V(p_+^\dagger p_- + p_-^\dagger p_+) +H_\text{\rm ph}\,,
\label{Ham-trans}
\eea
where $V_\pm=D_\mp^2 V_1 + D_\pm^2 V_2$ and $V=D_+D_- (V_1 - V_2)$.
The major outcome of this transformation is that the formerly diagonal interaction with the bath $H_{\rm IB}$, given by \Eq{HIB}, now develops the off-diagonal elements $V(p_+^\dagger p_- + p_-^\dagger p_+)$
which enable phonon-assisted transitions between the mixed qubit states. The transition rates can be evaluated via Fermi's golden rule (FGR)~\cite{MuljarovPRL05,MorreauPRB19}:
\be
\Gamma_- =  N_{R} \,\Gamma_\text{\rm ph}\,,\quad
\Gamma_+ =  (N_{R}+1)\, \Gamma_\text{\rm ph}\,,
\label{Eq:FRG}
\ee
where $N_{R}$ is the Bose function taken at the Rabi splitting $R$ and
\begin{equation}\label{FGR_gamma}
	\Gamma_\text{\rm ph} = \pi\sum_{\bf q}\left|D_+ D_- \left(\lambda_{{\bf q}, 1}-\lambda_{{\bf q}, 2}\right)\right|^2 \delta( v_s q-R)\,,
\end{equation}
according to the off-diagonal coupling in \Eq{Ham-trans}. Here the delta function expresses the energy conservation in real transitions, indicating that the energy difference between the mixed states should exactly match the energy of an emitted or absorbed phonon $\omega_q=v_s q$. The rate $\Gamma_\text{\rm ph}$ is evaluated in \App{App:FGR}, providing for an isotropic model of the QDs the explicit analytical result:
\be
\Gamma_\text{\rm ph} =\Gamma_0\left( 1-\frac{\sin(Rd / v_s)}{Rd /v_s } \right)\,,
\label{FGR}
\ee
where $\Gamma_0=D_+^2 D_-^2 R^3 (D_c - D_v)^2/(2\pi \rho_m v_s^5) e^{-l^2 R^2/v_s^2}$. The corresponding FGR calculation for an anisotropic model of the QDs is provided in \App{Sec-aniso}.

{ The FGR dephasing rates \Eq{FGR} are shown in \Fig{QDQD_Linebroadening}(b) as dashed lines, reproducing the main features of the exact calculation, but showing discrepancies (within $5\%$) due to multi-phonon processes not present in FGR.  The single-phonon transitions dominate at short distances as it is clear from the excellent agreement between the two results.}

\subsubsection{Physical interpretation of decoherence reduction}

The initial quadratic growth with distance, the oscillations, and the reduction of
$\Gamma_\pm$ at certain distances, seen in \Fig{QDQD_Linebroadening}(b), are all caused by the coherent properties of the phonon bath. According to \Eq{Ham-trans}, the phonon-assisted coupling between the mixed qubit states is given by $V_1-V_2$ which is proportional to $1-e^{i\textbf{q}\cdot \textbf{d}}$ [see \Eq{lambda}] and is vanishing at $\textbf{q}\cdot \textbf{d}=2\pi n$, where $n$ is an integer. This does not lead to a vanishing dephasing though, apart from $d=0$, owing to the 3D nature of the phonon momentum $\textbf{q}$ of the bath modes. However, as we show in \App{App:FGR}, in a 1D model of phonons with the same dispersion and same coupling, the dephasing rate \Eq{FGR} would modify to just
\be
\Gamma_\text{ph} =\Gamma_0 \left(\frac{v_s}{Rl}\right)^2 \sin^2 \left(\frac{Rd}{2v_s}\right)\,,
\label{Gamma1D}
\ee
strictly vanishing at $Rd/v_s=2\pi n$ for all $n$.
To understand this phenomenon in 1D, let us take the two-qubit state just before the event of phonon emission or absorption as a superposition $\alpha|1\rangle +\beta|2\rangle$ with some complex amplitudes $\alpha$ and $\beta$.
Since the qubits are entangled, they coherently emit or absorb the same phonon. This changes their phases (which is the source of pure dephasing) by $\varphi_1$ and $\varphi_2$, respectively, so that the two-qubit wave function becomes  $\alpha e^{i\varphi_1}|1\rangle +\beta e^{i\varphi_2}|2\rangle$, with $\varphi_2-\varphi_1=\pm qd$, according to \Eq{lambda} and energy conservation requiring $R=v_s q$. However, if the separation $d$ between the qubits is such that the phase difference is a multiple of $2\pi$, i.e. $Rd/v_s=2\pi n$ for an integer $n$, the resulting wave function only acquires a common phase factor $e^{i\varphi_1}$, which is not changing the state since there is no relative phase difference between qubit states. In other words, in order for the transition to occur between the initial and final states [e.g. between $|+\rangle$ and $|-\rangle$, see the inset in \Fig{QDQD_Linebroadening}(a)], which would result in a phonon-induced dephasing, a change of the two-qubit state is required, meaning that the interaction with a phonon must induce a relative phase shift, i.e. $Rd/v_s \neq 2\pi n$.

Note that in the case of e.g. nanowire-based QDs ~\cite{LindwallPRL07} or QDs in carbon nanotubes \cite{ArdizzonePRB15,JeantetNL17}, the phonon dispersion and coupling are altered when the dimensionality is reduced from a bulk system. Several branches of phonon modes arise due to the reduced dimensionality and phonon quantization which are not present in 3D systems. This leads to changes in the phonon dispersion and coupling matrix elements. As a result, there is a finite zero-phonon linewidth which is not observed in QDs coupled to bulk phonons, where the linewidth remains zero in the ideal case. Here, the ideal case corresponds to the condition $qd=2\pi n$, for which no broadening of the ZPL is observed, due to the phonon interactions not facilitating a change of state, making the system effectively equivalent to the independent boson model~\cite{Mahan00} in which there is no ZPL broadening.

\begin{figure}[t]
	\includegraphics[width=0.45\textwidth]{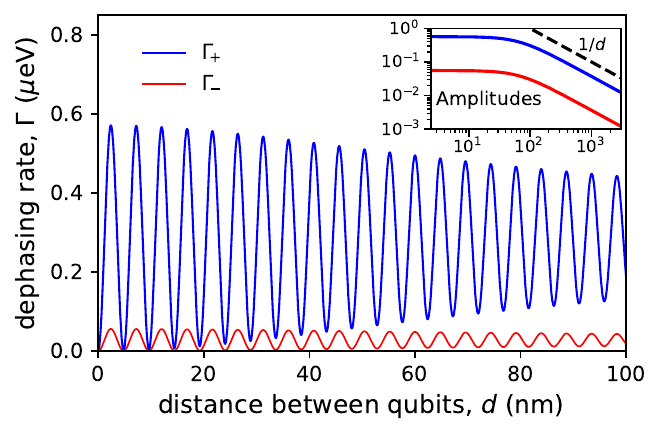}
	\caption{\label{FGRhighg}
		As \Fig{QDQD_Linebroadening}(c) but for $g=2\,$meV and FGR only (solid lines). Inset: Amplitude of the oscillations in the dephasing rates versus distance. }
\end{figure}

\subsubsection{Anisotropic QD qubits}

For 3D phonons and spherical QDs, the dephasing is absent only at $d=0$ and according to \Eq{FGR} and \Fig{QDQD_Linebroadening}(b) has minima around $Rd/v_s= 2\pi n+\pi/2$ ($n=1,2,\dots$). The $\pi/2$ phase shift compared to the 1D case and non-vanishing dephasing at the minima are due to phonons of energy $R$ that are absorbed or emitted at different angles $\theta$ to the QD separation vector ${\bf d}$ (\Fig{schematic}), resulting in a variation of their phase difference $\varphi_2-\varphi_1$ between the QDs.
However, the reduction of decoherence is enhanced in anisotropic QDs, playing the role of directional phonon emitters or absorbers. In fact, in oblate QDs separated along their short axis [\Fig{QDQD_Linebroadening}(c)], directional coupling of phonons along the short axis effectively makes the system 1D under certain conditions.

The dephasing rates of anisotropic QDs, calculated via FGR in \App{App:FGR} and having a compact analytical expression \Eq{FGRanis} in terms of the Faddeeva function, reproduce the main features of the exact calculation (with a relative difference below $7\%$), as seen in \Fig{QDQD_Linebroadening}(c).
In this case $l \gg l_\perp$,  where $l$ and $l_\perp$ are, respectively, the in-plane and perpendicular (along ${\bf d}$) exciton localization lengths, so that for $d \ll 2l^2 q$, where $q=R/v_s$, the dephasing rates vanish at $qd=2\pi n$, as it is clear from \Eq{F-F} in \App{sec-Faddeeva}.
If additionally  $ql \gg 1$, meaning that the relevant phonon wavelength is small enough to create a directional emission, the FGR dephasing rates reduce to \Eq{Gamma1D}. Under these conditions, the 3D system behaves as a 1D system, however, as the dot separation is increased, the 3D nature gradually returns. Furthermore, the 1D regime can be extended by increasing the anisotropy or increasing the energy $R$ of the dominant phonon modes which couple to the system.

In fact, the analogy with pure 1D phonons becomes striking for stronger coupled QDs ($g=2$\,meV) as shown in \Fig{FGRhighg}, where the shorter phonon wavelength involved in transitions provides fast oscillations versus $d$, allowing for minima at short distances with near-vanishing dephasing. With such coupling strengths, the aforementioned condition $ql \gg 1$ is met, having a value $ql=10$. The scaling of the oscillation amplitude with distance, given in the inset, demonstrates the quasi-1D behaviour (shown by constant amplitude) for $d\ll 2l^2 q \approx 148\,$nm. This is consistent with the first few minima in the main plot having visually very small dephasing rates before gradually returning to the 3D regime as the dot separation increases. For this directional emission of phonons, the phonon Rayleigh length, given by $d_R = l^2 q/2 \approx 37$\,nm, estimates how far the phonons can propagate as a focused beam, maintaining 1D-like behavior. Beyond this distance, the system gradually transitions back to 3D.  For 1D behavior to persist, the condition on the qubit separation then becomes $ d\ll 4 d_R$, where  $d_R$ serves as a reasonable upper limit for ensuring the system remains in the 1D regime.

\begin{figure}[t]
	\centering
	\includegraphics[width=0.45\textwidth]{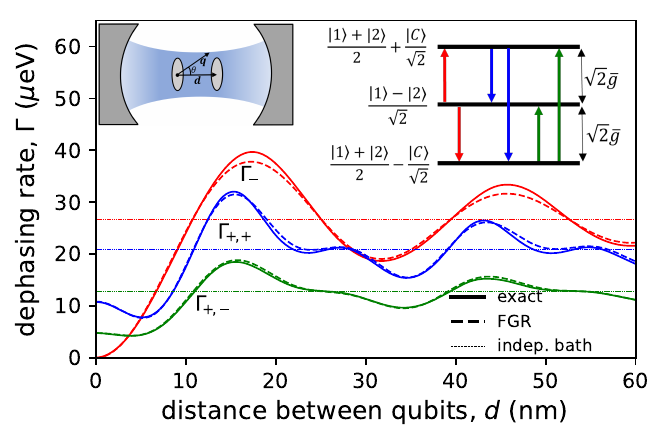}
	\caption{Dephasing rates $\Gamma_{+,\pm}$ and $ \Gamma_{-}$ of the mixed states as a function of $d$. Calculated exactly (solid lines) and via FGR (dashed lines) for cavity-mediated coupled anisotropic qubits (left inset) with interaction strength $g_1=g_2=\bar{g}=0.5\,$meV, and zero detuning. The dephasing rates for independent phonon baths are shown by thin dashed lines. Right inset: Energy level diagram for the mixed qubit-cavity states $|+,\pm\rangle=(|1\rangle+|2\rangle)/2\pm|C\rangle/\sqrt{2}$ and $|-\rangle=(|1\rangle-|2\rangle)/\sqrt{2}$, with real phonon-assisted transitions (red, blue and green arrows). Other parameters as in \Fig{QDQD_Linebroadening}(c). }
	\label{QDQDCAV_Linebroadening}
\end{figure}

\subsection{Cavity-mediated coupled QD qubits}

In {\em Case B} of QDs indirectly coupled via a cavity, the dephasing is also controlled by bath coherence, though in a more complex scenario. For zero detuning ($\Omega_1=\Omega_2=\Omega_C$) and equal QD-cavity couplings ($g_1=g_2=\bar{g}$), the resulting three coupled states, $|+,\pm\rangle=(|1\rangle+|2\rangle)/2\pm|C\rangle/\sqrt{2}$ and  $|-\rangle=(|1\rangle-|2\rangle)/\sqrt{2}$ require a triexponential fit of $P_{11}(t)$ to extract the dephasing rates $\Gamma_j=-\Im\, \omega_j$, which are shown in \Fig{QDQDCAV_Linebroadening} across a range of distances, see \App{App:Fit} for details of the fit.  We observe oscillations versus distance $d$, different from those of directly coupled QDs (\Fig{QDQD_Linebroadening}) since there are two periods contributing to the dephasing rates $\Gamma_{+,\pm}$ of the states $|+,\pm\rangle$. This is due to the involvement of transitions at two distinct frequencies, as seen in the right inset, with one twice the other (the general case of a non-zero detuning with three different frequencies involved is considered in \App{App:FGR2}). Since the dephasing rate $\Gamma_-$ of the state $|-\rangle$ involves transitions to the two other levels with equal Rabi splitting and thus the interacting phonons have almost the same energy, only one period is observed in the oscillations of $\Gamma_-$, analogous to {\em Case A}, with vanishing dephasing rate at $d=0$. In general, $\Gamma_{+,+}$ (consisting of two downwards transitions) will always be greater than $\Gamma_{+,-}$ (two upwards transitions), simply because of spontaneous phonon emission. Furthermore, whether $\Gamma_-$ or $\Gamma_{+,+}$ is the largest on average depends on the coupling strength chosen. If the Rabi splitting ($\sqrt{2}\bar{g}$) for the transitions contributing to $\Gamma_-$ is closer to the peak in the phonon spectral density than the energy ($2\sqrt{2}\bar{g}$) of the distant-level transitions included in $\Gamma_{+,-}$, then $\Gamma_-$ is the largest dephasing rate.

Due to the nature of the mixed QD-cavity states, the exciton-phonon matrix elements contributing to FGR are now proportional to $V_1\pm V_2$, with $+$ ($-$) corresponding to the transitions between distant (neighboring) levels, see \App{App:FGR2} for details. Since transitions between distant levels contribute to $\Gamma_{+,\pm}$ and thus involve $V_1 + V_2$, there is a non-vanishing contribution even at $d=0$. This is because the states involved in such transitions both have a cavity contribution, and as cavity does not couple to phonons, the reduction of the dephasing rate to zero is not observed. However, these transition have typically lower impact on decoherence due to the larger phonon energy involved, as discussed in more detail in \App{App:FGR2}.

For two-qubit or multiple-qubit systems, the gate fidelity, as described in \Onlinecite{Nielsen2010-jf}, is a function of the actual density matrix, quantifying how well a quantum gate performs compared to the ideal operation. The ideal gate represents a unitary operation (i.e. without phonons), and the interaction of the system with phonons typically causes decoherence, disrupting the ideal unitary evolution and reducing fidelity. As we have demonstrated, the optical polarization, which corresponds to the off-diagonal elements of the full density matrix, can be reduced or even eliminated in certain cases. The influence of the dephasing on the diagonal elements of the density matrix does not exceed the one on the related off-diagonal elements, by construction. Therefore, in the special case when decoherence is eliminated, the diagonal elements are also unaffected, and the density matrix represents the ideal case, having maximum gate fidelity.

\section{Conclusion}

In conclusion, we have presented an asymptotically exact solution for the linear optical response of a system of two coupled qubits interacting with a shared bath, using semiconductor quantum dots coupled to 3D acoustic phonons as illustration. While coupling to the bath causes decoherence, we have shown that the coherence of the bath itself can be exploited to reduce such decoherence. By controlling the distance between the qubits in relation to the wavelength of the interacting bath modes, it is possible to minimize decoherence. We find that for a 1D bath, decoherence can be eliminated entirely, a case which can also be approached for anisotropic qubits in a 3D bath. This concept can be generalized to multiple-qubit systems.

\section*{Acknowledgments}
L.H. acknowledges support from the EPSRC under
grant no. EP/T517951/1. L.S. acknowledges support
from the EPSRC under grant no. EP/R513003/1.

\appendix

\section{Exciton-phonon coupling elements and phonon spectral density}
\label{App:Coupling}

Throughout this work, we consider semiconductor QDs as candidates for qubits, using typical InGaAs parameters outlined in \cite{MuljarovPRL04,MuljarovPRL05}. At low temperatures, the exciton-phonon interaction is primarily governed by the deformation potential coupling with longitudinal acoustic phonons. Assuming that the phonon parameters within the QDs closely resemble those of the surrounding material, and further assuming that the acoustic phonons exhibit linear dispersion, $\omega_q = v_s q $, where $q=|{\bf q}|$ and $v_s$ is the sound velocity in the material, the exciton-phonon matrix coupling element for an exciton in qubit $j=1,2$ is given by
\begin{equation} \label{lambda_D}
	\lambda_{\textbf{q},j} = \frac{\sqrt{q}\,\mathcal{D}_j(\textbf{q})}{\sqrt{2\rho_m v_s \mathcal{V}}}\,,
\end{equation}
where $\rho_m$ is the mass density of the material, $\mathcal{V}$ is the sample volume, and
\begin{equation}\label{form-factor}
	\mathcal{D}_j(\textbf{q})= \int d\textbf{r}_{e} \int d\textbf{r}_{h}  \abs{\Psi_{X,j}(\textbf{r}_{e},\textbf{r}_{h})}^2 \left(D_c e^{i\textbf{q}\cdot \textbf{r}_{e}} -D_v e^{i \textbf{q}\cdot \textbf{r}_{h}} \right)
\end{equation}
is the coupling form-factor \cite{MuljarovPRL04,MuljarovPRL05}, with $D_{c(v)}$ being the deformation potential of the conduction (valence) band. Assuming a factorizable form of the exciton wave functions, $\Psi_{X,j}(\textbf{r}_{e},\textbf{r}_{h})= \psi_{e,j}(\textbf{r}_{e}) \psi_{h,j}(\textbf{r}_{h})$, where $\psi_{e(h),j}(\textbf{r})$ is the confined electron (hole) ground state wave function in QD $j$, the form-factor simplifies to
\begin{equation}\label{deformationpotential}
	\mathcal{D}_j(\textbf{q}) = \int d\textbf{r} \left[ D_c\abs{\psi_{e,j}(\textbf{r})}^2 - D_v \abs{\psi_{h,j}(\textbf{r})}^2\right] e^{i\textbf{q} \cdot \textbf{r} }\,.
\end{equation}

\subsection{Isotropic quantum dots (QDs)}
\label{iso_QDs}
Choosing spherically symmetric parabolic confinement potentials, the ground-state wave functions of the carriers take Gaussian form, which in the simpler case of equal electron and hole confinement lengths, $l_{e,j} = l_{h,j} = l_j$, is given by
\begin{equation} \label{psi_j}
	\psi_j(\textbf{r}) = \frac{1}{\pi^{3/4} l_j^{3/2}} \exp{-\frac{ (\textbf{r} - \textbf{d}_j)^2}{2l_j^2}}\,,
\end{equation}
where $\textbf{d}_j$ is the coordinate of the center of QD $j$.  Substituting \Eq{psi_j} into \Eq{deformationpotential}, performing the integration over the whole space and substituting the result into \Eq{lambda_D}, we obtain
\begin{equation}\label{lambda_j}
	\lambda_{\textbf{q},j} = \sqrt{q} \lambda_0 \exp{-l_j^2 q^2/4} e^{i \textbf{q}\cdot\textbf{d}_j}\,,
\end{equation}
where
\begin{equation}
	\lambda_0 = \frac{D_c - D_v}{\sqrt{2\rho_m v_s \mathcal{V}}}\,.
\end{equation}
Choosing the first QD located at the origin ($\textbf{d}_1=0$) we have $\textbf{d}_2=\textbf{d}$, where $\textbf{d}$ is the distance vector between the QDs.
Converting the summation over ${\bf q}$ to an integration, $\sum_{\bf q} \rightarrow \frac{\mathcal{V}}{(2\pi)^3} \int d {\bf q}$, and using spherical coordinates, the spectral density $J_{jj'}(\omega)$ defined by \Eq{spectraldens} takes the form
\begin{equation}
	\begin{split}
		J_{jj'}(\omega) = &\frac{J_0 v^4_s}{2} \int_0^\infty dq\, q^3 \exp{-q^2 l^2} \delta(\omega-v_s q) \\
		&\times\int_0^\pi d\theta \sin\theta
		\begin{cases}
			1 & j=j' \\
			\exp{iqd\cos\theta} & j < j' \\
			\exp{-iqd\cos\theta} & j > j'\,,
		\end{cases}
	\end{split}
\end{equation}
where
\begin{equation}\label{J0}
	J_0 = \frac{(D_c - D_v)^2}{4\pi^2 \rho_m v_s^5}\,,
\end{equation}
$d=|{\bf d}|$, and $l^2 = (l_j^2 + l_{j'}^2)/4$ (for brevity omitting the indices $j$ and $j'$ in the new length $l$ introduced). Performing the integration over the polar angle $\theta$, we arrive at
\begin{equation}\label{spectraldens_isotropic}
	J_{jj'}(\omega)= J_0 \, \omega^3 \exp{-\frac{ \omega^2 l^2}{v_s ^2}}
	\times
	\begin{cases}
		1 & j=j' \\
		{\rm sinc}\left(\frac{\omega d}{v_s}\right) & j \neq j'\,,
	\end{cases}
\end{equation}
where ${\rm sinc}(x)=\sin(x)/x$.

\subsection{Anisotropic QDs}
\label{Aniso_QDs}
For anisotropic QDs with in-plane confinement length $l_j$ and perpendicular confinement length $l_{j,\perp}$, the Gaussian ground-state wave functions \Eq{psi_j} are modified to
\bea
\psi_j(x,y,z) &=& \frac{1}{\pi^{3/4} l_j l_{\perp,j}^{1/2}} \exp{-\frac{ ({x -d_ {x,j}})^2+ ({y -d_ {y,j}})^2}{2l_j^2}}
\nonumber\\
&&\times \exp{-\frac{ ({z -d_ {z,j}})^2}{2l_{\perp,j} ^2}}\,,
\label{psi_anisotropic}
\eea
where we have again taken the case of identical electron and hole localization lengths, $l_{e,j} = l_{h,j} = l_j$ and $l_{\perp,e,j} = l_{\perp,h,j} = l_{\perp,j}$, and
used the components $(d_ {x,j},d_ {y,j},d_ {z,j})$ of the vector ${\bf d}_j$. Following the same procedure as for isotropic qubits, we obtain
\begin{equation}\label{lambda_anisotropic_qxqyqz}
	\lambda_{\textbf{q},j} = \sqrt{q} \lambda_0 \exp{-l_j^2 (q_x ^2 + q_y^2)/4-l_{\perp,j}^2 q_z^2/4} e^{i \textbf{q}\cdot\textbf{d}_j}\,.
\end{equation}
The above equation is assuming that both QDs have the same anisotropy axis (along $z$).
Assuming further that the centers of the QDs lie on the $z$-axis, so that $d_ {x,j}=d_ {y,j}=0$ and $d_ {z,j}=d_j$, we find in spherical coordinates
\bea\label{lambda_anisotropic}
\lambda_{\textbf{q},j} &=&  \sqrt{q} \lambda_0 \exp\left(-q^2(l_j^2 \sin^2(\theta)+l_{\perp,j}^2 \cos^2(\theta))/4\right.\nonumber\\&&+i q d_j \cos(\theta)\big{)}\,,
\eea
using $q_x = q\sin(\theta)\cos(\phi)$, $q_y = q\sin(\theta)\sin(\phi)$, and $q_z = q\cos(\theta)$.
The spectral density is then given by
\begin{equation}\label{spectraldens_anisotropic}
	\begin{split}
		J_{jj'}(\omega) & = \frac{J_0 v^4_s}{2} \int_0^\infty dq\, q^3 \delta(\omega-v_s q)
		\\
		&\times
		\int_0^\pi d\theta  \sin(\theta) \exp{-q^2 l^2 \sin^2(\theta)-q^2 l_\perp^2 \cos^2(\theta)} \\
		&\times
		\begin{cases}
			1 & j=j' \\
			\exp{iqd\cos(\theta)} & j < j' \\
			\exp{-iqd\cos(\theta)} & j > j'\,,
		\end{cases}
	\end{split}
\end{equation}
where $l^2 = (l_j^2 + l_{j'}^2)/4$ and $l_\perp^2 =  (l_{\perp,j}^2 + l_{\perp,j'}^2)/4$.
Performing the integration over the polar angle $\theta$, we obtain
\begin{equation}
	\label{J-anisotropic}
	J_{jj'}(\omega)\!=\! J_0 \, \omega^3 e^{-q^2 l_\perp^2}
	\times
	\begin{cases}
		\!F\left(0,q\sqrt{l_\perp^2 - l^2} \right) & \!j=j' \\
		\!F\left(\frac{d}{2\sqrt{l_\perp^2 - l^2}},q\sqrt{l_\perp^2 - l^2} \right) & \!j \neq j'
	\end{cases}
\end{equation}
with $q=\omega/v_s$, where
\be
\label{F-def}
F(\alpha, \beta)= \frac{\sqrt{\pi}}{4\beta} \left[ e^{-2i\alpha\beta}w(\alpha-i\beta)-e^{2i\alpha\beta}w(\alpha+i\beta) \right]\,,
\ee
and $w(z)$ is the Faddeeva function. Note that \Eq{J-anisotropic} is valid for both $l_\perp>l$ and $l_\perp<l$, and in the isotropic case $l_\perp=l$ simplifies to \Eq{spectraldens_isotropic}, as shown in \Sec{sec-Faddeeva} below.

\subsection{Faddeeva function and some properties of $F(\alpha, \beta)$ }
\label{sec-Faddeeva}

The Faddeeva function $w(z)$ is defined as
\be
w(z)= \frac{2}{\sqrt{\pi}}\int_0^\infty e^{2izt}e^{-t^2} dt\,,
\label{Faddeeva}
\ee
for any complex number $z$. Physically, it has the meaning of a convolution of Gaussian and complex Lorentzian functions. In fact, for Im\,$z>0$, \Eq{Faddeeva} is equivalent to
\be
w(z)= \frac{i}{\pi} \int_{-\infty}^\infty \frac{e^{-t^2}}{z-t}\, dt\,.
\label{Faddeeva2}
\ee
The Faddeeva function has the properties
\be
\label{Faddeeva-prop}
w(-z)=2e^{-z^2}-w(z) \rmand
\left[w(z)\right]^\ast =w(-z^\ast)\,,
\ee
and is linked to the error function $\erf (z)$ by
\be
w(z)=e^{-z^2} \left[1+\erf(iz)\right]\,,
\ee
where
\be
\erf (z)=\frac{2}{\sqrt{\pi}} \int_0^z e^{-t^2} dt\,.
\ee
It can also be expressed in terms of the Dawson function $D(z)$ as
\be
w(z)= e^{-z^2} +\frac{2i}{\sqrt{\pi}} D(z)
\ee
where
\be
D(z)= \int_0^\infty e^{-t^2}\sin(2zt) dt = e^{-z^2} \int_0^z e^{t^2} dt\,.
\ee
Clearly, all three functions, $w(z)$, $\erf(iz)$, and $D(z)$, are equivalent in the sense that they can be expressed by each other. Analytically, the error function has an advantage that it is an entire function, so that $[\erf(z)]^\ast = \erf(z^\ast)$, in addition to being an odd function, $\erf(-z)=-\erf(z)$. However, numerically, the Faddeeva function (as well the Dawson function) is generally more accurate and stable, since the error function $\erf(z)$ diverges at large imaginary values of $z$, but the Faddeeva and Dawson functions do not.

The function $F(\alpha, \beta)$, introduced in \Eq{F-def} can also be written as
\be
\label{F-def2}
F(\alpha, \beta)= \frac{1}{2} \int_{-1}^1  e^{\beta^2 (1-x^2)} e^{2i\alpha\beta x} dx\,,
\ee
reflecting the integration over the polar angle in \Eq{spectraldens_anisotropic}.
It has the properties
\be
F(\alpha, \beta)=F(-\alpha, \beta)=F(\alpha,-\beta)=F^\ast(\alpha, \beta)\,,
\ee
which are easy to show using the definition \Eq{F-def2}, but can be obtained also from the analytic form \Eq{F-def} and the properties of the Faddeeva function, \Eq{Faddeeva-prop}.

For $\alpha=0$, corresponding to $d=0$ in \Eq{J-anisotropic}, one has
\be
F(0,\beta)=\frac{\sqrt{\pi}}{4}\,\frac{w(-i\beta) - w(i\beta)}{\beta} = \frac{\sqrt{\pi}}{2}\,e^{\beta^2}\,\frac{ \erf (\beta)}{\beta}
\ee
and in the limit $\beta\to0$, corresponding to isotropic dots ($l_\perp=l$) or zero-frequency ($q=0$),
\be
\lim_{\beta\to 0} F(0,\beta) =1 \,,
\ee
so that \Eq{J-anisotropic} simplifies to \Eq{spectraldens_isotropic}.

In the isotropic limit ($l_\perp=l$), $
\beta \to0$ and $\alpha = qd/(2\beta)\to\infty$, and we obtain from \Eq{F-def}
\bea
\lim_{\beta\to 0} F\left(\frac{qd}{2\beta},\beta\right)&=& -\frac{\sqrt{\pi}}{4} \left(e^{iqd}-e^{-iqd}\right)
\lim_{\beta\to 0} \frac{1}{\beta} w\left(\frac{qd}{2\beta}\right)
\nonumber\\
&=& {\rm sinc}\,(qd)\,,
\eea
using
\be
\lim_{z\to \infty} zw(z) = \lim_{z\to \infty} \frac{1}{\sqrt{\pi}} \int_0^\infty \hspace*{-3mm} e^{iz'}\exp{-\frac{z'^2}{2z^2}} dz' = \frac{i}{\sqrt{\pi}}
\label{zw}
\ee	
with $z'/z$ being real, as it follows  from the definition \Eq{Faddeeva}, again, in agreement with \Eq{spectraldens_isotropic}.

Let us finally consider the limit of a strong anisotropy, $l\gg l_\perp$ which is used at the end of \Sec{Sec-aniso} below. In this limit,
\be
\alpha = \frac{d}{2\sqrt{l_\perp^2 - l^2}} \approx \frac{-id}{2l} \rmand  \beta=q\sqrt{l_\perp^2 - l^2} \approx iql \,,
\ee
with $2\alpha\beta = qd$. Under the condition that $|\alpha| \ll |\beta|$ (equivalent to $d\ll 2l^2q$) one can then obtain from \Eq{F-def}
\bea
&&F(0,\beta)-F(\alpha,\beta)\approx
\nonumber\\
&\approx& -\frac{\sqrt{\pi}}{4iql} \left[ (1-e^{iqd})w(-ql)- (1-e^{-iqd})w(ql)\right]
\nonumber\\
&=& \frac{\sqrt{\pi}}{4 iql} \left[4 w(ql)\sin^2\frac{qd}{2} - (1-e^{iqd})e^{-q^2l^2}\right],
\label{F-F}
\eea
using \Eq{Faddeeva-prop}. Clearly, this function vanishes if $\sin(qd/2)=0$. In the case of $ql \gg 1$ this simplifies to just
\be
F(0,\beta)-F(\alpha,\beta)\approx
\frac{\sqrt{\pi}}{iql} w(ql)\sin^2\frac{qd}{2} \approx \frac{1}{q^2l^2} \sin^2\frac{qd}{2}\,,
\label{FFanis}
\ee
using the limit \Eq{zw}.

\section{Explicit form of the memory kernal and the cumulant elements}
\label{App:Cumulants}

Using the cumulant element $K_{jj'}(s)$ and \Eq{Kinim} allows us to provide an explicit expression for the memory kernel \Eq{GtensorLN}:
\bea
\mathcal{G}_{i_L \dots i_1 l} &= &M_{i_1 l}  \exp\left\{ \xi_{l}\xi_{l}K_{11}(0) + \eta_{l}\eta_{l}K_{22}(0) \right.
\nonumber\\  &&
+ 2[\xi_{i_1}\xi_{l}K_{11}(1) + \eta_{i_1}\eta_{l}K_{22}(1)
\nonumber\\  &&
+ (\xi_{i_1}\eta_{l} + \eta_{i_1}\xi_{l})K_{12}(1)]+ \dots
\nonumber\\
&&  + 2[\xi_{i_L}\xi_{l}K_{11}(L) + \eta_{i_L}\eta_{l}K_{22}(L)
\nonumber\\  &&
+ (\xi_{i_L}\eta_{l} + \eta_{i_L}\xi_{l})K_{12}(L)]\}\,.
\label{Gtensor}
\eea
The above expression is valid for the shared bath. For independent baths, the relevant modification of the tensor $\mathcal{G}_{i_L \dots i_1 l}$ consists of setting in \Eq{Gtensor} all the mixed terms to zero, since $K_{12}(s)=0$ according to \Eq{KinimIndependent}.

To calculate the cumulant elements \Eq{Kjj'}, we use the fact that $K_{jj'}(|n-m|)$ depends on the difference $|n-m|$ only, and not on both time steps $n$ and $m$ individually. We therefore can find them recursively using the values of the cumulant functions \Eq{Cjj} on the time grid, starting from
\be
K_{jj'}(0) = C_{jj'}(\Delta t)\,.
\ee
The remaining $s > 0$ cumulant elements are found recursively via
\bea
K_{jj'}(s) &=& \frac{1}{2} \Biggl[C_{jj'}\big((s+1)\Delta t\big) - (s+1)K_{jj'}(0)
\nonumber\\
&&\hspace{6mm}-2\sum_{h=1}^{s-1} (s+1 -h) K_{jj'}(h) \Biggr].
\label{C_k}
\eea

\section{Choosing the time step in the Trotter decomposition approach}
\label{App:Cutoff}

The energy separation $R$ between the mixed states determines the timescale
\begin{equation}
	\tau_{0} = \frac{2\pi}{R}\,,
\end{equation}
which is the period of the corresponding Rabi rotations. In the discretization used in the $L$N approach described in \Sec{Sec-LN}, this timescale should be much larger than the time step $\Delta t$ of discretization, $\Delta t \ll \tau_{0}$. In the cavity-coupled two-qubit system, $R$ can take three different values, and the above condition should be fulfilled for all of them. For example, at zero detuning ($\Omega_1=\Omega_2=\Omega_C$), the same coupling to the cavity ($g_1=g_2=\bar{g}$), and no dipolar coupling ($g=0$), the largest energy separation is evaluated as $R=2\sqrt{2}\bar{g}$, see the inset in \Fig{QDQDCAV_Linebroadening}. In the case of the dipolar coupled QDs without a cavity, there are only two mixed states and therefore only one Rabi splitting, evaluated to $R=2g$ at zero detuning ($\Omega_1=\Omega_2$). At the same time, the polaron timescale $\tau_{\rm IB}$ is given by~\cite{MorreauPRB19}
\begin{equation}
	\tau_{\rm IB} \approx \frac{\pi \sqrt{l^2 + l_\perp^2}} {v_s}
\end{equation}
for anisotropic QDs with in-plane ($l$) and perpendicular ($l_\perp$) Gaussian lengths. The polaron timescale characterizes the time to form or disperse a polaron cloud following the creation or destruction of an exciton in a QD. The selected time step $\Delta t$ must be large enough such that for a given number of $L+1$ time steps within the memory kernel, the resulting memory time of $(L+1)\Delta t$ is larger than $\tau_\text{\rm IB}$. Specifically, the total time considered via the time steps must cover the dynamics of the cumulant $\mathcal{K}_{i_n i_m}$ defined in Eq.\eqref{Kinim}, which is dependent on the cumulant elements $C_{11}$, $C_{12}$, and $C_{22}$, with the full temporal evolution defined in \Eq{Cjj}.

\begin{figure}[t]
	\centering
	\includegraphics[scale=0.8]{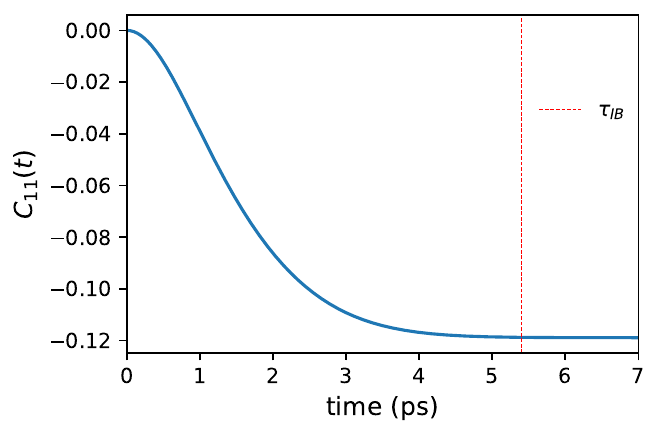}
	\caption{\label{fig:CumulantElementC11} Temporal evolution of $C_{11}(t)$ (blue line) and the phonon memory time $\tau_\text{\rm IB}$ (vertical red dashed line). The parameters are as in \Fig{QDQD_Linebroadening}(c) resulting in $\tau_\text{\rm IB}=5.39\,$ps.
	}
\end{figure}

\begin{figure}[t]
	\centering
	\includegraphics[scale=0.8]{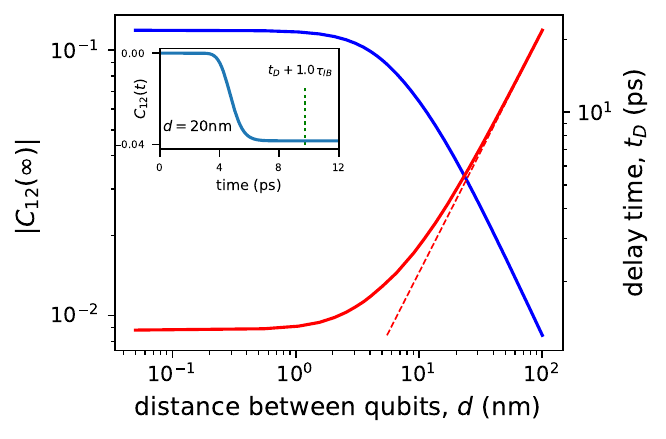}
	\caption{\label{fig:CumulantElementC12} The asymptotic value of $|C_{12}(t)|$ (blue line, left axis) and the delay time $t_D$ (red line, right axis) as functions of the interdot distance $d$, with the red dashed line being the estimate $d/v_s$ of the time taken for a phonon to travel between the QDs. The inset shows the temporal evolution of $B(t)$ at $d=20$nm, demonstrating the delay time, decay and saturation at a minimum value. The green vertical dashed line in the inset shows the full memory time considered. The parameters used are the same as in \Fig{fig:CumulantElementC11}.
	}
\end{figure}

Focusing on the cumulant element $C_{11}(t)$, we see from \Fig{fig:CumulantElementC11} that $(L+1)\Delta t \geq \tau_\text{\rm IB}$ is in fact sufficient to fully cover the dynamics due to this element $C_{11}(t)$. In practice, however, one should perform a convergence test for the chosen parameters, to ensure the full memory time is taken into account.  In the case of identical QDs, $C_{11}(t)=C_{22}(t)$, otherwise the larger $\tau_\text{\rm IB}$ of the two QDs should be used. However since both QD excitons couple to the same phonons, there are extra cumulant elements $K_{12}(s)$ which depend on the distance $d$ separating the QDs. The effect of this distance dependence is the introduction of a delay time before $C_{12}(t)$ starts to change, this can be seen in the inset of \Fig{fig:CumulantElementC12}. Physically this delay time is due to the time it takes a phonon to travel between the QDs, which is approximately $d/v_s$. For consistency, we define in the calculations the delay time $t_D$ to be the time at which the change of $C_{12}(t)$ is equal to a half of its minimum value, i.e. $C_{12}(t_D)=C_{12}(\infty)/2$. The values of $t_D$ are shown in \Fig{fig:CumulantElementC12} (red curve) as function of the interdot distance, along with its rough estimate $d/v_s$ (red dashed line) working well at large distances.

The presence of the delay time $t_D$ in the cumulant function $C_{12}(t)$ implies that the time step in discretization must be increased to cover the full memory time of $C_{12}(t)$, so the condition
$\Delta t = \tau_\text{\rm IB}/(L+1)$ suitable for a QD-cavity system~\cite{MorreauPRB19} is no longer sufficient for distant coupled QDs with increasing QD separation $d$. We therefore modify this condition to  \begin{equation}\label{timestep}
	\Delta t = \frac{t_D +  \tau_\text{\rm IB}}{L+1}\,,
\end{equation}
which takes the delay time into account, thus covering the memory time for all cumulant elements. The green vertical dashed line in the inset of \Fig{fig:CumulantElementC12} demonstrates that all changes of the cumulant functions, $C_{ij}(t)$, are covered over the memory time $\Delta t (L+1)$.

As the memory time increases due to the increase in delay time with increasing $d$, the accuracy of the
calculation decreases for a given $L$ due to the increase in time step. As seen from the inset, $C_{12}(t)$ saturates at a minimum value $C_{12}(\infty)$, and the blue line in \Fig{fig:CumulantElementC12} shows the decrease of $|C_{12}(\infty)|$ as $d$ increases, implying that $C_{12}(t) \rightarrow 0$ as $d\rightarrow \infty$. This means in the limit of $d\rightarrow \infty$, the full shared phonon bath calculation becomes equivalent to the independent bath case, which is naturally expected, whereby the result is now independent of the distance between the QDs and therefore $\Delta t = 1.0\,\tau_\text{\rm IB}/(L+1)$ is again sufficient since there is no delay time through the $K_{12}(s)$ cumulant elements.

\section{Fermi's golden rule -- dipolar coupled qubits}
\label{App:FGR}
\begin{figure}[t]
	\centering
	\subfloat[\centering ] {{\includegraphics[width=0.25\textwidth]{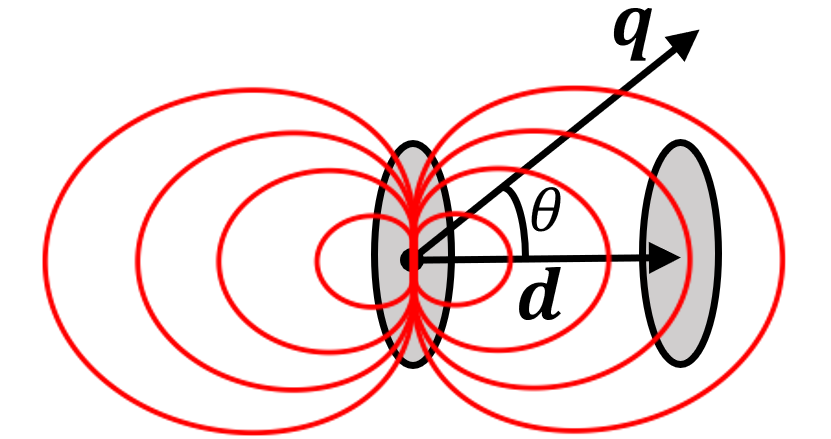} }}%
	\subfloat[\centering ] {{\includegraphics[width=0.25\textwidth]{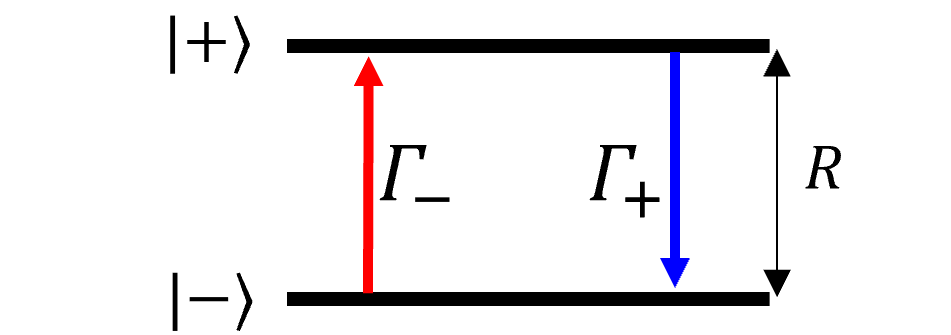} }}%
	\caption{\label{QDQD_energydiagram}
		(a) Schematic of the system for a pair of dipole-dipole coupled anisotropic QDs separated by distance $d$ and a phonon with the wave vector $\textbf{q}$ emitted or absorbed at an angle $\theta$ (for clarity the dipole-dipole interaction is shown only for the left QD acting on the right QD). (b)
		Nonzero-detuning energy level diagram for the mixed states $\ket{\pm}$ composed from the basis states $\ket{1}$ and $\ket{2}$ of isolated QDs. Red and blue arrows show phonon-assisted transitions between the mixed states, resulting in the line broadening $\Gamma_-$ and $\Gamma_+$  of the lower and upper states,  respectively.
	}
\end{figure}

In this appendix, we apply instead of the canonical transformation \Eq{canon} used in the main text a unitary transformation to the full Hamiltonian \Eq{Ham} of the system in {\em Case A}, considering two directly coupled QDs without cavity. Following this transformation, we use FGR to calculate the phonon-assisted transition rates between the hybrid QD states, as illustrated in \Fig{QDQD_energydiagram}, and consequently the dephasing rates of the linear polarization.

Let us consider the full Hamiltonian $H=H_{0} + H_\text{\rm IB}$, defined in \Eqs{H0}{HIB} with the cavity coupling $g_1=g_2=0$. In the basis of pure QD states, $|1\rangle$ and $|2\rangle$, $H_{0}$ has the following matrix form
\begin{equation}
	H_{0} =
	\begin{pmatrix}
		\Omega_1 & g \\ g & \Omega_2
	\end{pmatrix}.
\end{equation}
This matrix can be diagonalized by a unitary transformation $S^\dagger H_{0}S=\Lambda$, where
\begin{equation}
	S=S^{-1}=S^\dagger =
	\begin{pmatrix}
		D_- & D_+ \\ D_+ & -D_-
	\end{pmatrix},
\end{equation}
with $D_\pm$ given by \Eq{Dpm} and
\begin{equation}
	\Lambda = \begin{pmatrix}
		\Omega_+ & 0 \\ 0 & \Omega_-
	\end{pmatrix}
\end{equation}
being a diagonal matrix of the eigenvalues \Eq{ompm}.

Applying this transformation to the full Hamiltonian, we obtain
\bea
\tilde{H}&=&S^\dagger H S =
\begin{pmatrix}
	\Omega_+ & 0 \\ 0 & \Omega_-
\end{pmatrix}
\nonumber\\
&&+ \begin{pmatrix}
	D_- & D_+ \\ D_+ & -D_-
\end{pmatrix} \begin{pmatrix}
	V_1 & 0 \\ 0 & V_2
\end{pmatrix}
\begin{pmatrix}
	D_- & D_+ \\ D_+ & -D_-
\end{pmatrix}
+H_\text{ph}\mathbb{1}
\nonumber\\
&=&
\begin{pmatrix}
	\Omega_+ + V_+& V \\ V & \Omega_- + V_-
\end{pmatrix}
+ H_\text{ph}\mathbb{1}\,,
\label{H-trans}
\eea
with $V_\pm$ defined in the main text, and $\mathbb{1}$ being the $2\times2$ identity matrix. The main outcome of this transformation is the off-diagonal coupling to phonons given by $V=D_+D_- (V_1 - V_2)$. This coupling is responsible for the phonon-assisted transitions between the hybrid (or mixed) states and ultimately for the long-time dephasing of the optical polarization.

\subsection{Isotropic QDs}
\label{Sec-iso}

Here we evaluate the rate $\Gamma_\text{ph}$  in FGR \Eq{Eq:FRG} for isotropic QDs, substituting \Eq{lambda_j} into \Eq{FGR_gamma}, converting the summation over ${\bf q}$ to an integration and further expressing the integration in spherical coordinates, we find
\bea
\Gamma_\text{ph} &=& \frac{D_+^2 D_-^2 (D_c -D_v)^2}{8\pi \rho_m v_s}
\int_0^\infty dq \,q^3 e^{-q^2l^2}
\\
&\times&
\int_0^\pi d\theta \, \sin(\theta) ( 2- e^{iqd \cos\theta} - e^{-iqd \cos\theta})\delta(v_s q -R)\,,
\nonumber
\eea
where $l^2 = (l_1^2 + l_{2}^2)/4$ (for identical QDs $l_1=l_2=l\sqrt{2}$).
Integrating over $\theta$, we obtain
\begin{equation}\label{gamma_isotropic}
	\Gamma_\text{ph} = \frac{D_+^2 D_-^2 (D_c - D_v)^2}{2\pi \rho_m v_s^5} R^3 e^{-\frac{ R^2 l^2 }{v_s^2}}
	\left[ 1-{\rm sinc}\left(\frac{Rd}{v_s}\right) \right]\,.
\end{equation}
In the case of zero detuning, $R=2g$ and $D_+ = D_- = 1/\sqrt{2}$. In the limit of $d\rightarrow \infty$, ${\rm sinc}(Rd/v_s) \rightarrow 0$, so that $\Gamma_\text{ph}$ becomes independent of $d$. In the limit of $d \rightarrow 0$, ${\rm sinc}(x) \approx 1-x^2/6$, leading to a $d^2$ dependence at small distances and vanishing dephasing rates at $d=0$.

Let us note also that for a 1D phonon bath which is for example the case of a QD embedded in a quantum wire, the latter providing a 2D quantum confinement of phonon modes, \Eq{FGR_gamma} would give instead, for the same coupling matrix element \Eqs{lambda_D}{form-factor} and the linear phonon dispersion $\omega=v_s q$, the following dependence on the Rabi splitting $R$ and interdot distance $d$:
\begin{equation}
	\label{FGR_1D_isotropic_qdqd}
	\Gamma_\text{ph} \propto R \, e^{-\frac{R^2 l^2_\perp }{v_s^2}} \sin^2\left(\frac{Rd}{2v_s}\right),
\end{equation}
where $l_\perp$ is the Gaussian length of the electron and hole confinement in the direction of the phonon propagation.

\subsection{Anisotropic QDs}
\label{Sec-aniso}	

Performing a similar calculation for anisotropic QDs, we find, after substituting the exciton-phonon coupling element \Eq{lambda_anisotropic} into \Eq{FGR_gamma}:
\bea
\Gamma_\text{ph} &=& \frac{D_+^2 D_-^2 (D_c -D_v)^2}{8\pi \rho_m v_s}
\int_0^\infty dq \,q^3
\\
&&\times
\int_0^\pi d\theta \, \sin(\theta) e^{-q^2l^2\sin^2\theta} e^{-q^2l_\perp^2\cos^2\theta}
\nonumber\\
&&\times
\left( 2- e^{iqd \cos\theta} - e^{-iqd \cos\theta}\right)\delta(v_s q -R)\,,
\nonumber
\eea
where $l^2 = (l_1^2 + l_2^2)/4$ and $l_\perp^2 = (l_{\perp,1}^2 + l_{\perp,2}^2)/4$ (for identical QDs $l_1=l_2=l\sqrt{2}$ and $l_{\perp,1}=l_{\perp,2}=l_\perp \sqrt{2}$).
Performing the integration, we obtain
\bea
\Gamma_\text{ph} &=&\frac{D_+^2 D_-^2 (D_c - D_v)^2}{2\pi \rho_m v_s^5} R^3
e^{- q^2 l^2_\perp}
\Biggl[F\left(0,q\sqrt{l_\perp^2 - l^2} \right)
\nonumber\\
&&- F\left(\frac{d}{2\sqrt{l_\perp^2 - l^2}},q\sqrt{l_\perp^2 - l^2} \right)\Biggr],
\label{FGRanis}
\eea
where $q=R/v_s$ and the function $F(\alpha,\beta)$ is given by \Eq{F-def}.

For strongly anisotropic QDs with $l\gg l_\perp$, $R l/v_s \gg 1$ (small phonon wavelength) and $d \ll 2l^2 q$ ($|\alpha| \ll |\beta|$), we find using \Eq{FFanis}
\begin{equation}\label{FGR_1D_anisotropic_qdqd}
	\Gamma_\text{ph} =\frac{D_+^2 D_-^2 (D_c - D_v)^2}{2\pi \rho_m v_s^3}\, \frac{R}{l^2}\,
	e^{-\frac{R^2 l^2_\perp }{v_s^2}} \sin^2\left(\frac{Rd}{2v_s}\right),
\end{equation}
which has the same dependence on the distance $d$ and the Rabi splitting $R$ as in the model of 1D phonons \Eq{FGR_1D_isotropic_qdqd}.

\section{Fermi's golden rule -- cavity-mediated coupled qubits}
\label{App:FGR2}
\begin{figure}[t]
	\centering
	\subfloat[\centering ] {{\includegraphics[width=0.23\textwidth]{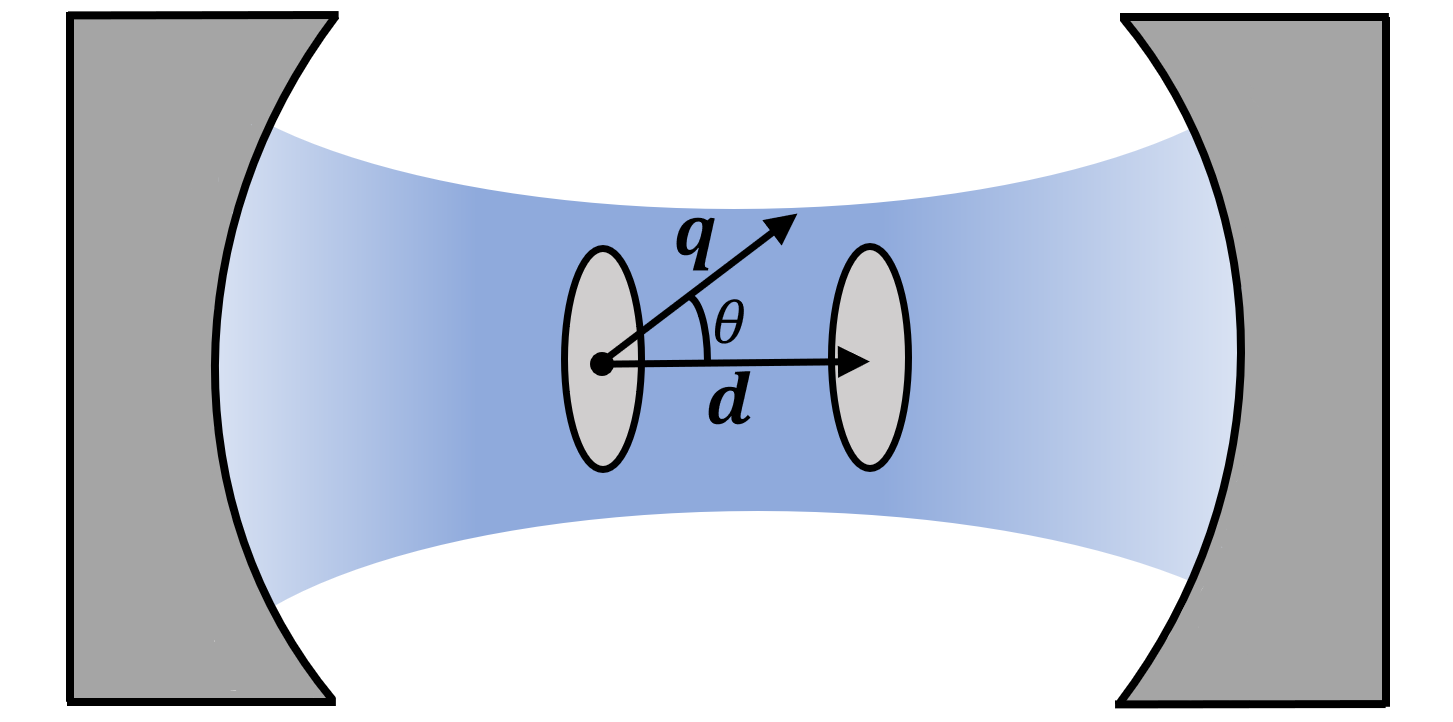} }}%
	\subfloat[\centering ] {{\includegraphics[width=0.23\textwidth]{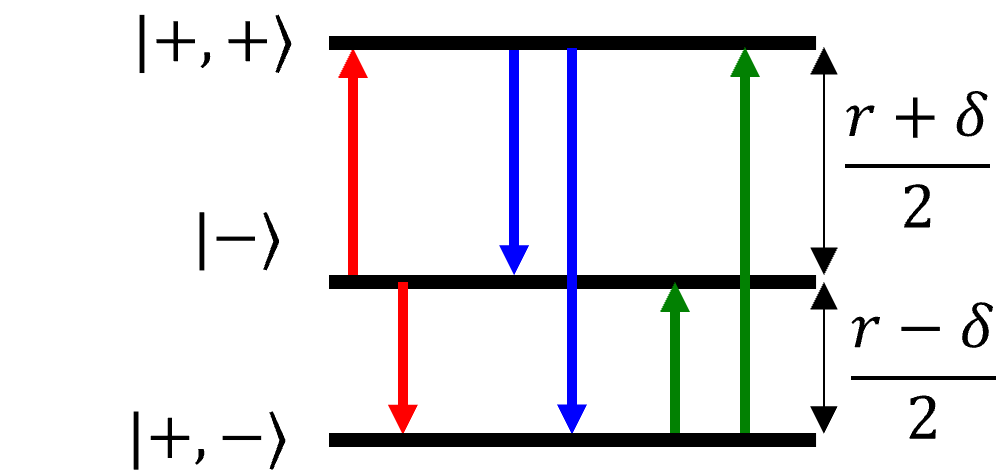} }}%
	\caption{\label{QDQDCAV_energydiagram} (a) Schematic of the system for a pair of anisotropic QDs separated by distance $d$, each interacting independently with the cavity mode
		and a phonon with the wave vector $\textbf{q}$ emitted or absorbed at an angle $\theta$.
		(b) The $g_1=g_2=\bar{g}$ and nonzero-detuning ($\delta\neq0$) energy level diagram  for the mixed states $\ket{+,\pm}$ and $\ket{-}$. These mixed states are composed from the basis states $\ket{1}$, $\ket{2}$, and $\ket{C}$ of the isolated QDs and the cavity with a single photon. The transitions indicated by the red, blue and green arrows result in the line broadening (dephasing rates) of the central, upper, and lower states, denoted by $\Gamma_-$, $\Gamma_{+,+}$, and $\Gamma_{+,-}$, respectively.
	}
\end{figure}

Let us now focus on the other special case ({\em Case B}) of no direct dipolar coupling of two QD qubits, i.e. $g=0$, but an indirect coupling mediated by their interaction with a common cavity mode with the coupling constants $g_1$ and $g_2$. Reducing the full basis to pure QD states, $|1\rangle$ and $|2\rangle$, and the single-photon cavity state $|C\rangle$, which is sufficient for the linear polarization, the Hamiltonian of the cavity-mediated system takes the form
\begin{equation}
	\label{H-QDcavity}
	H=H_0 + V_1\ket{1}\bra{1} + V_2\ket{2}\bra{2}+ H_\text{ph}\,,
\end{equation}
where
\bea
H_0&=& \; \Omega_1 \ket{1}\bra{1} + \Omega_2 \ket{2}\bra{2} + \Omega_C\ket{C}\bra{C}
\label{H_0_qdqdcav}\\
&&+ g_1(\ket{1}\bra{C} + \ket{C}\bra{1}) + g_2(\ket{2}\bra{C} + \ket{C}\bra{2})\,,
\nonumber
\eea
and  $H_\text{ph}$ and $V_i$ are given by \Eq{HIB}. We apply a transformation diagonalizing $H_0$ as $S^\dagger H_0 S = \Lambda$, so the full Hamiltonian transforms to
\bea
\tilde{H} &=& S^\dagger H S
\\
&=&\mathbb{1} H_\text{ph}+S^\dagger\begin{pmatrix}
	\Omega_1 & 0 & g_1 \\ 0 & \Omega_2 & g_2 \\ g_1 & g_2 & \Omega_C
\end{pmatrix} S  + S^\dagger\begin{pmatrix}
	V_1 & 0 & 0 \\ 0 & V_2 & 0 \\ 0 & 0 & 0
\end{pmatrix} S \,,
\nonumber
\eea
where $\mathbb{1}$ is the $3\times3$ identity matrix. In general, $H_0$ is diagonalized numerically, providing the mixed state energy eigenvalues $\Lambda_j$. The transformation of the exciton-phonon coupling generates off-diagonal elements responsible for phonon-assisted transition between mixed QD-cavity states which we account for below using FGR.

Focusing on the analytically solvable case of zero detuning between the QD qubit states, $\Omega_1=\Omega_2=\Omega$ (e.g. for identical qubits), and the same coupling of both qubits to the cavity, $g_1=g_2=\bar{g}$, the transformation matrix has the following explicit form
\be
S=\begin{pmatrix}
	d_- & \frac{1}{\sqrt{2}} & d_+ \\ d_- & -\frac{1}{\sqrt{2}} & d_+ \\ \sqrt{2} d_+ & 0 & -\sqrt{2} d_-
\end{pmatrix}\,,
\ee	
where
\be
d_\pm = \frac{1}{2} \sqrt{1 \pm \frac{\delta}{r}}
\ee
with
\begin{equation}
	r= \sqrt{\delta^2 + 8\bar{g}^2}
	\rmand
	\delta=\Omega_C- \Omega\,,
\end{equation}
the latter being the cavity-QD detuning. The Hamiltonian \Eq{H-QDcavity} then transforms to
\bea
\tilde{H} &=& S^\dagger H S = \mathbb{1} H_\text{ph}
\label{H-QDcavity-trans}\\
\small
&&+\begin{pmatrix}
	\Omega+\frac{\delta +r}{2}+ U_+ d^2_- & \frac{U_- d_-}{\sqrt{2}} & U_+ d_+ d_- \\
	\frac{U_- d_-}{\sqrt{2}} & \Omega + \frac{U_+}{2}& \frac{U_- d_+}{\sqrt{2}}  \\
	U_+ d_+ d_- & \frac{U_- d_+}{\sqrt{2}}  & \Omega+\frac{\delta -r}{2}+ U_+ d^2_+
\end{pmatrix} \,,
\nonumber
\eea
where $U_\pm = V_1 \pm V_2$. By applying this transformation, we go from the $\ket{1}$, $\ket{2}$, $\ket{C}$ basis to the mixed state basis
\bea
\ket{+,\pm}&=&d_\mp(\ket{1}+\ket{2})\pm\sqrt{2}d_\mp\ket{C}\,,
\nonumber\\
\ket{-}&=& (\ket{1}-\ket{2})/\sqrt{2}\,,
\eea
analogous to that in the polariton transformation of a qubit-cavity system outlined in \cite{MorreauPRB19}. Figure \ref{QDQDCAV_energydiagram} illustrates the level structure of the mixed states for nonzero detuning ($\delta\neq0$) and the phonon-assisted transitions due to the off-diagonal elements in \Eq{H-QDcavity-trans}. The rates of these transitions are estimated below via FGR, similar to \Sec{Sec:FGR}:
\begin{equation}\label{transitions}
	\Gamma_{\uparrow,\pm} = {\cal N}_R \Gamma_{\text{ph}, \pm} \rmand
	\Gamma_{\downarrow,\pm} =({\cal N}_R+1) \Gamma_{\text{ph}, \pm}\,,
\end{equation}
respectively, for the upwards and downwards transitions,
where
\begin{equation}\label{Gamma_ph_pm}
	\Gamma_{\text{ph}, \pm} = \pi \sum_q \abs{c_0 (\lambda_{\textbf{q},1} \pm \lambda_{\textbf{q},2}) }^2 \delta(\omega - R)\,.
\end{equation}
There are six possible transitions corresponding to the six off-diagonal matrix elements in \Eq{H-QDcavity-trans}. ${\cal N}_R$ is the Bose distribution function taken at the energy $R$, which is the separation of the energy levels of the mixed states involved in the transition and $c_0$ is the corresponding factor. These energy levels are given by $\Lambda_{+,\pm}=\Omega+(\delta \pm r)/2$ and $\Lambda_-=\Omega$, according to \Eq{H-QDcavity-trans}. In particular, for $ |+,-\rangle \leftrightarrow |-\rangle $ transitions, $R=(r-\delta)/2$ and $c_0=d_+/\sqrt{2}$; for $|-\rangle \leftrightarrow |+,+\rangle$ transitions,  $R=(r+\delta)/2$ and $c_0=d_-/\sqrt{2}$; finally, for $|+,-\rangle \leftrightarrow |+,+\rangle$ transitions,  $R=r$ and $c_0=d_+ d_-$. Note that for the phonon-assisted transitions between the neighboring levels ($|+,-\rangle \leftrightarrow |-\rangle$ and $|-\rangle \leftrightarrow |+,+\rangle$), the exciton-phonon coupling matrix elements $\lambda_{\textbf{q},j}$ contribute to \Eq{Gamma_ph_pm} as a difference due to $U_-$, thus giving $\Gamma_{\text{ph}, -}$, and for transitions between the distant levels ($|+,-\rangle \leftrightarrow |+,+\rangle$) as a sum due to $U_+$, thus giving $\Gamma_{\text{ph}, +}$, see \Eq{H-QDcavity-trans}.

Using the same procedure as in \App{App:FGR}, we evaluate the transition rates \Eq{Gamma_ph_pm} for identical isotropic and anisotropic QD qubits. For isotropic dots, \Eq{Gamma_ph_pm} yields
\begin{equation}\label{gamma_isotropic2}
	\Gamma_{\text{ph}, \pm} = \frac{c_0^2 (D_c - D_v)^2}{2\pi \rho_m v_s^5} R^3 e^{-\frac{ R^2 l^2 }{v_s^2}}
	\left[ 1\pm{\rm sinc}\left(\frac{Rd}{v_s}\right) \right],
\end{equation}
where the difference to \Eq{gamma_isotropic} are the constant factors, the energy distance $R$, and most importantly the presence of the $\pm$ sign before the sinc function, differentiating the neighboring ($-$) and the distant ($+$) level transitions. Note that the contribution of the distant level transitions to the decoherence is typically less significant due to the factor $e^{- R^2 l^2 /v_s^2}$ in which $R^2$ is four times larger (for zero detuning) than for the neighboring level transitions. Similarly, for anisotropic QDs we find
\bea
\Gamma_{\text{ph}, \pm} &=&\frac{c_0^2 (D_c - D_v)^2}{2\pi \rho_m v_s^5} R^3
e^{- q^2 l^2_\perp}
\Biggl[F\left(0,q\sqrt{l_\perp^2 - l^2} \right)
\nonumber\\
&&\pm F\left(\frac{d}{4\sqrt{l_\perp^2 - l^2}},q\sqrt{l_\perp^2 - l^2} \right)\Biggr]
\label{gamma_anisotropic2}
\eea
with $q=R/v_s$.

Using \Eq{gamma_isotropic2} or \Eq{gamma_anisotropic2} in combination with \Eq{transitions}, the contribution to the line broadening for a specific phonon-assisted transition can be found. The line broadening  $\Gamma_{+,\pm}$ and $\Gamma_-$ of the mixed states is the sum of the broadening by the two available transitions.

\section{Triexponential fit of the polarization for cavity-mediated coupled QD qubits}
\label{App:Fit}
\begin{figure}[t]
	\centering
	\includegraphics[scale=0.8]{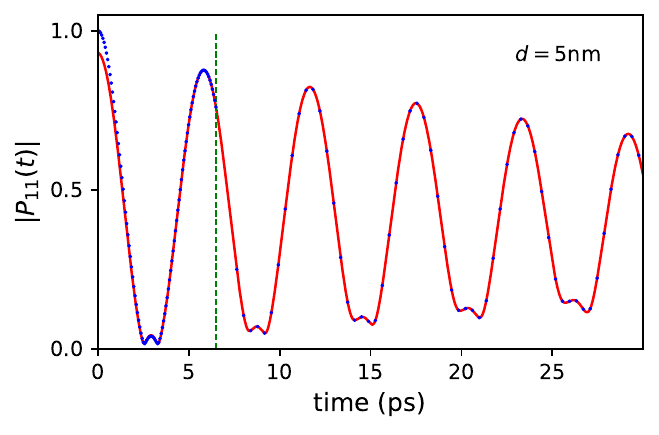}
	\caption{
		Linear optical polarization $|P_{11}(t)|$ (blue dots) and its complex tri-exponential fit (red lines) for cavity mediated coupled anisotropic QD qubits at zero detuning, separated by a distance $d=5\,$nm, with excitation and measurement in QD $1$. The parameters are as in \Fig{QDQD_Linebroadening}(c).
	}
	\label{QDQDCAV_P1}	
\end{figure}

We show in \Fig{QDQDCAV_P1} the optical linear polarization $|P_{11}(t)|$ for {\em Case B} in the main text. The linear polarization (blue dots) for cavity-coupled QD qubits again starts from unity due to the excitation and measurement of the same QD state and has the temporal oscillations now at three frequencies due to addition of a cavity mode. We apply a complex triexponential fit (red curve) of the form $\sum_j {C}_j e^{ -i\omega_j t }$, extracting the complex amplitudes ${C}_j$, energies $\Re\,\omega_j$, and dephasing rates $\Gamma_j=-\Im\, \omega_j$ of the phonon-dressed mixed states. The fit is applied after the phonon-memory cut-off (dashed green vertical line), beyond the polaron cloud formation time. The dephasing rates are then extracted across a range of distances, providing Fig. 4 of the main text.

\section{Extrapolation of fit parameters}
\label{App:Extrapolation}
As detailed in the main text, \Figs{QDQD_Linebroadening}{QDQDCAV_Linebroadening} are created by calculating the linear optical polarization for a given number of neighbors ($L$), then applying a fit to the long time data and extracting the fit parameters. The parameters corresponding to the line broadening, $\Gamma(L)$, are extracted across a range of neighbors, and the convergence of $\Gamma(L)$ to the exact ($L=\infty$) value is assumed to follow a power law model, given by:
\begin{equation}
	\Gamma(L) = \Gamma(\infty) + CL^{-\beta}.
\label{GamL}
\end{equation}

\begin{figure}[t]
	\centering
	\includegraphics[scale=0.8]{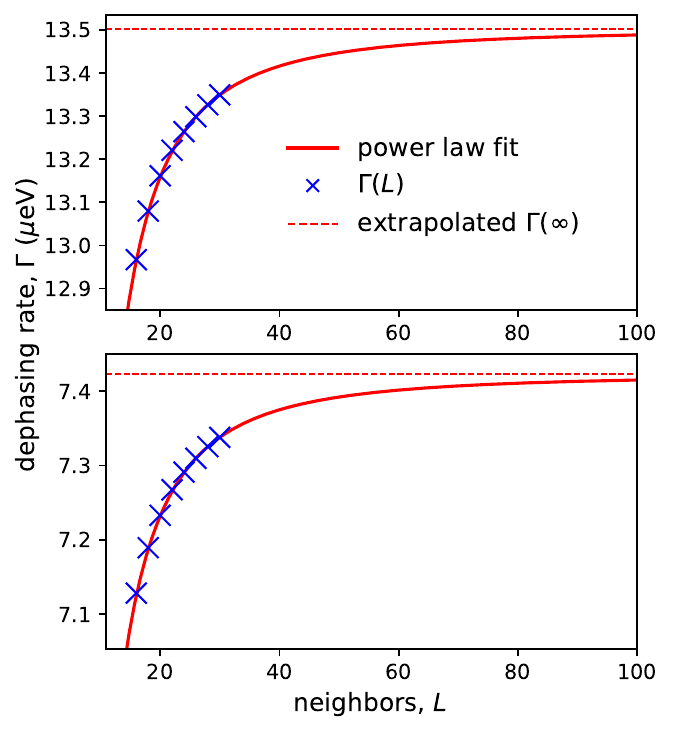}
	\caption{
		Power law fit applied to the $\Gamma_+(L)$ ($\Gamma_-(L)$) values across a range of neighbors, $L$, for $d=5\,$nm is shown in the upper (lower) figure. The blue crosses are the extracted $\Gamma(L)$ values, the red curve is the power law model with $\beta=2$, and the red horizontal dashed line is the estimated value of $\Gamma(\infty)$. The parameters are as in \Fig{QDQD_Linebroadening}(b).
	}
	\label{PL_fit}	
\end{figure}
\vspace{-0.15em}

Figure~\ref{PL_fit} shows the $\Gamma(L)$ calculated values (blue crosses) for directly coupled QD qubits treated in {\em Case A}, with the power law model applied (red curve), and the extrapolated $\Gamma(\infty)$ is shown as a red dashed line. The value of $\Gamma(\infty)$ is estimated for the eight valued of $\Gamma(L)$ shown in \Fig{PL_fit}, by minimizing the root mean square deviation from the power law
\Eq{GamL} for $\beta=2$.

\FloatBarrier 

%

\end{document}